\documentclass[
  twoside,
  twocolumn,
  preprintnumbers,
  floatfix,
  prb,
  superscriptaddress
]{revtex4-1}

\usepackage{graphicx}
\usepackage{epstopdf}
\usepackage{tabularx}
\usepackage{times}
\usepackage{amsmath}
\usepackage{dcolumn}
\usepackage{pifont}
\usepackage{hyperref}

\newcommand{\Ti}{\text{Ti}}
\newcommand{\Fe}{\text{Fe}}
\newcommand{\Cu}{\text{Cu}}
\renewcommand{\O}{\text{O}}
\newcommand{\PT}{\text{PbTiO$_3$}}
\newcommand{\MB}{\text{$M_B$}}
\newcommand{\fM}{\text{$f_{M}$}}

\newcommand{\Vab}{\text{$V_{ab}$}}
\newcommand{\Vc}{\text{$V_c$}}

\newcommand{\MBVab}{\MB-\Vab}

\newcommand{\MBVcA}{\MB-\text{$V_{c1}$}}
\newcommand{\MBVcB}{\MB-\text{$V_{c2}$}}
\newcommand{\K}[0]{\text{K}}
\newcommand{\eV}[0]{\text{eV}}
\newcommand{\THz}[0]{\text{THz}}
\newcommand{\cm}[0]{\text{cm}}
\newcommand{\s}[0]{\text{s}}
\renewcommand{\deg}{\ensuremath{^\circ}}

\newcommand{\sect}[1]{Sect.~\ref{#1}}
\newcommand{\fig}[1]{Fig.~\ref{#1}}
\newcommand{\eq}[1]{Eq.~(\ref{#1})}
\newcommand{\tab}[1]{Table~\ref{#1}}

\renewcommand{\tensor}[1]{\ensuremath\boldsymbol{\mathrm{#1}}}
\newcommand{\tp}[1]{#1^{\mathrm{T}}}
\renewcommand{\vec}[1]{\ensuremath\boldsymbol{#1}}
\renewcommand{\epsilon}[0]{\varepsilon}

\begin{document}

\title{
  Formation and switching of  defect dipoles in acceptor doped lead titanate: \newline
  A kinetic model based on first-principles calculations
}
\date{\today}

\author{Paul Erhart}
\email{erhart@chalmers.se}
\affiliation{
  Chalmers University of Technology,
  Department of Applied Physics,
  SE-412 96 Gothenburg, Sweden
}
\author{Petra Tr\"askelin}
\affiliation{
  University of Gothenburg,
  Department of Physics,
  SE-412 96 Gothenburg, Sweden
}
\author{Karsten Albe}
\affiliation{
  Institut f\"ur Materialwissenschaft,
  Technische Universit\"at Darmstadt,
  D-64287 Darmstadt, Germany
}

\begin{abstract}
The formation and field-induced switching of defect dipoles in acceptor doped lead titanate is described by a kinetic model representing an extension of the well established Arlt-Neumann model [Ferroelectrics {\bf 76}, 303 (1987)]. Energy barriers for defect association and reorientation of oxygen vacancy-dopant (Cu and Fe) complexes are obtained from first-principles calculations and serve as input data for the kinetic coefficients in the rate equation model. The numerical solution of the model describes the time evolution of the oxygen vacancy distribution at different temperatures and dopant concentrations in the presence or absence of an alternating external field. We predict the characteristic time scale for the alignment of all defect dipoles with the spontanenous polarization of the surrounding matrix. In this state the defect dipoles act as obstacles for domain wall motion and contribute to the experimentally observed aging. Under cycling conditions the fully aligned configuration is perturbed and a dynamic equilibrium is established with defect dipoles in parallel and anti-parallel orientation relative to the spontaneous polarization. This process can be related to the deaging behavior of piezoelectric ceramics.
\end{abstract}


\maketitle

\section{Introduction}
\label{sect:intro}

Aging phenomena, namely the gradual change of physical properties with time, are observed in almost all ferroelectrics. \cite{Ple56, IkeUed67, CarHar78, Tak82, ArlNeu88, LohNeuArl90, WarDimPik95, AfaPetPro01, ZhaRen05, ZhaErdRen08, MorDam08, GenGlaHir09, ZhaRen10} In some acceptor doped barium titanate ($\rm BaTiO_3$) and lead zirconate titanate (PZT) ceramics aging goes along with an increasing shift of the hysteresis along the axis of the electrical field giving rise to an internal bias field. \cite{TagStoCol01} In the past, several plausible models have been developed to intepret the occurence of bias fields and aging phenomena in ferrocelectrics in terms of domain splitting \cite{IkeUed67}, space-charge formation, \cite{Tak82} electronic charge trapping, \cite{WarDimPik95, AfaPetPro01}, ionic drift \cite{MorDam08} and reorientation of defect dipoles. \cite{ArlNeu88, Ren04, ShiGriChe07, JiaMiUrb08}

In acceptor (``hard'') doped ferroelectrics transition metals usually substitute the $B$-site (Ti or Zr in PZT) and tend to bind strongly to oxygen vacancies. These acceptor center-oxygen vacancy associates form electric and elastic defect dipoles such as charged $(\Fe'_{\text{Zr},\Ti}$-$V^{\bullet \bullet}_\O)^{\bullet}$  or $(\Cu''_{\text{Zr},\Ti}$-$V_\O^{\bullet\bullet})^\times$, \cite{MesEicKlo05, ErhEicTra07} which contribute to the overall polarization in a ferroelectric compound \cite{WarDimPik95, PoyCha99a, MesEicDin04, MesEicKlo05, ErhEicTra07, BooSmiChe07, EicErhTra08, MarEls11} and can be aligned either parallel, anti-parallel, or perpendicular to the polarization of the surrounding material as shown schematically in \fig{fig:confs}.
 
In the parelectric state, defect dipoles of different orientation are energetically equivalent, whereas they have a preferred orientation in a polar matrix. Arlt and Neumann \cite{NeuArl87, ArlNeu88} have attributed the occurence of internal bias fields to the switching of defect dipoles and described the transient orientation of dipoles by a kinetic model. As quantitative data on the energy landscape for these defect dipoles was unavailable at the time they relied on a very simple electrostatic estimate of the energy difference.
\footnote{
  In Ref.~\onlinecite{ArlNeu88} an alternative estimate for the energy differences between different defect dipole alignments is given based on dipolar interaction, which leads larger energy difference that are closer to the ones obtained by first-principles calculations. These values were, however, not employed in said reference to actually model aging.
}
The energetic asymmetry between the parallel and anti-parallel dipoles obtained in this fashion for BaTiO$_3$ was about 30\,meV and thus much smaller than the energy differences calculated more recently by first-principles methods for PbTiO$_3$, \cite{PoyCha99a, MesEicDin04, ErhEicTra07} which revealed that the energetic asymmetry is actually  as large as the barriers for oxygen migration. Only recently, Marton and Els\"asser \cite{MarEls11} showed that in Fe-doped lead titanate the barrier for reorientation sensitively depends on the position of the migrating oxygen vacancy with respect to the iron atom and the surrounding spontaneous ferroelectric polarization. During fast field cycling, the defect-dipoles are expected not to change orientation, because the characteristic rate for oxygen jumps around the acceptor center should be lower than the domain switching process.  Experimentally, Zhang {\it et al.} \cite{ZhaErdRen08} followed the dynamics of $(\text{Mn}_{\Ti}$-$V_\O)^\times$ dipoles in barium titanate by electron paramagnetic resonance studies and found support for the so-called "defect symmetry principle", which assumes that non-switching defect dipoles impose a restoring force for reversible domain switching. \cite{Ren04} Jakes {\it et al.} could show that in $\rm Fe^{3+}$ doped PZT defect dipoles are not preferentially located at domain walls but within the domains. \cite{JakErdEic11} Morozov {\it et. al.} studied aging-deaging process in hard PZT ceramics using the harmonic analysis of polarization response under switching conditions and concluded that two or more mechanisms are responsible for domain stabilization. \cite{MorDam08} Activation energies of about 0.6\,eV were attributed to short-range charge hopping, which could be due to local reorientation of microdipoles. 

Since the switching dynamics of defect dipoles depends on the electric and thermal energy provided to change polarization direction, the contribution of dipole reorientation can only be reliably assessed if realistic numbers for the migration and association energies are available, which allow to quantitatively model the switching dynamics of defect dipoles in a comprehensive way.  

The objective of the present work is to develop a kinetic model that captures the formation of defect dipoles as well as their reorientation both in the absence and presence of electric fields.
\footnote{
  Since the direct contribution of the electric field to the energy landscape is very small (see \sect{sect:results_ac}), within our model the results for non-oscillating (DC) fields are virtually identical to the situation without any external field.
}
Cu and Fe-doped lead titanate are considered as representative examples and the energy landscape for oxygen vacancy migration in these materials is obtained using first-principles calculations. We consider both free oxygen vacancies and oxygen vacancies associated with Fe or Cu. Starting from a statistical distribution the majority of oxygen vacancies is initially unbound. Over time vacancies are captured by impurity atoms and subsequently converted into the lowest energy configuration, which corresponds to a defect dipole that is aligned parallel to the macroscopic polarization ($M_B-V_{c1}$ in \fig{fig:confs}). While the exact time scales for these processes are dependent on dopant type, concentration, and temperature our results demonstrate that the ground state is reached within seconds at temperature slightly above room temperature and thus that already the pistine material can be considered as ``aged''. In the presence of an oscillating external field our model predicts that a gradual reorientation of defect dipoles leads to a dynamic equilibrium, in which the parallel and anti-parallel configurations occur with equal probability. This is in accord with the experimental observation of deaging by the application of AC fields. \cite{GraSuvKun06, GlaGenKun12}

This paper is organized as follows. First, we describe the kinetic model and discuss its features. This is followed in \sect{sect:dft} by a description of the first-principles calculations that were carried to determine the model parameters. In \sect{sect:results} we apply the kinetic model to study vacancy redistribution as a function of temperature and impurity concentration both in the absence and presence of an oscillating external electric field. The implications of the present findings for aging and fatigue are discussed in \sect{sect:discussion} and conclusions are summarized in \sect{sect:conclusions}.

\section{Kinetic model}
\label{sect:kinmodel}

\begin{figure*}
  \centering
\includegraphics[width=0.98\linewidth]{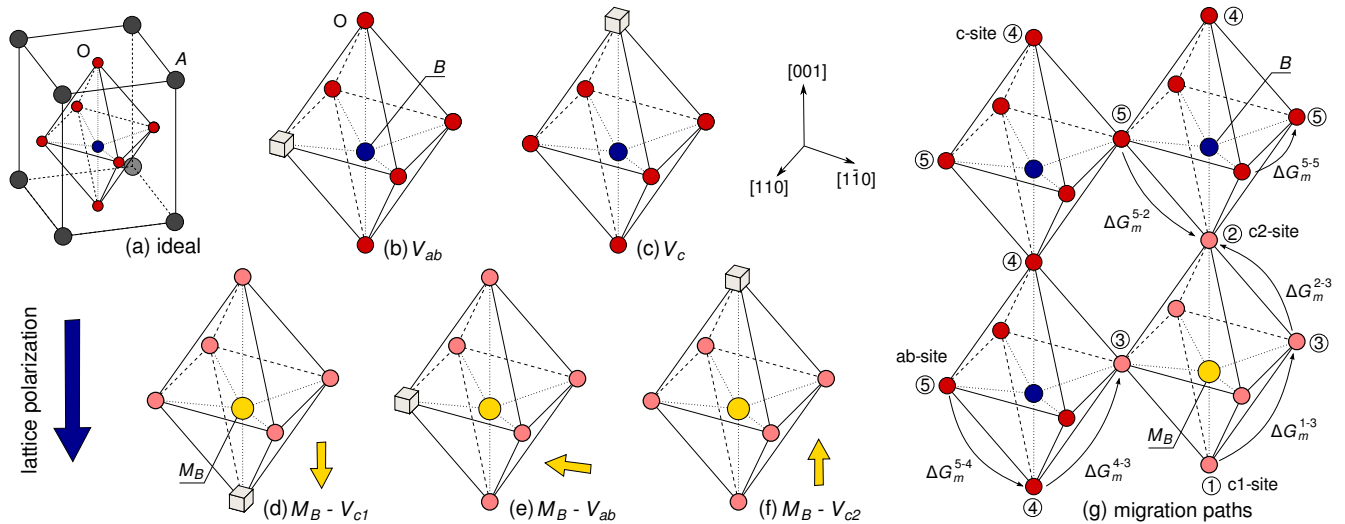}
  \caption{
    (a) Unit cell of the ideal tetragonal perovskite ($AB\O_3$) lattice.
    (b-c) Free oxygen vacancies.
    (d-f) Complexed oxygen vacancies on nearest-neighbor sites of $B$-site impurities.
    The dark blue arrow on the left indicates the direction of the lattice polarization whereas the smaller yellow arrows represent the orientation of the defect dipoles.
    (g) Schematic representation of the possible paths for oxygen vacancy migration. Some migration barriers, $\Delta G_m^{i-j}$, are exemplarily indicated. In (b-g) Only the $B$ and O sites are shown. Oxygen vacancies jump along the vertices of the $B$O$_6$ octahedron. Blue and yellow circles represent native and impurity atoms on $B$ sites, respectively. The dark and light red circles indicate oxygen atoms in the first nearest neighbor shell of native and impurity atoms, respectively.
  }
  \label{fig:migration_paths}
  \label{fig:confs}
\end{figure*}

In this section we formulate a kinetic model that describes the redistribution of oxygen vacancies between different types of sites as a function of time. It captures the temperature, impurity concentration and frequency dependence of this process within a mean-field approximation. Figure~\ref{fig:confs} provides an overview of the different types of oxygen vacancies that are taken into account by this model.

In general, the temporal variation of the concentration of vacancies of type $i$ can be described by a rate equation
\begin{align}
  \frac{\partial c_i}{\partial t}
  &=
  - \sum_j \Phi_{ij} \mathrm{K}_{ij} c_i
  + \sum_j \Phi_{ji} \mathrm{K}_{ji} c_j
  \label{eq:crate},
\end{align}
where the first term on the right hand side accounts for the ``loss'' of vacancies of type $i$ while the second term describes the ``gain'' due to vacancy jumps from sites of type $j$ to sites of type $i$. At typical device operation temperatures near 300\,K the creation or annihilation of vacancies at surfaces or interfaces is negligible and the total concentration of vacancies can be assumed as constant, 
\begin{align}
  \sum_i c_i = c_{tot}.
  \label{eq:ctot}
\end{align}
The rate at which vacancies of type $i$ jump onto sites of type $j$ is given by
\begin{align*}
  {\mathrm K}_{ij} &=
  \nu^i_0 \exp\left(-\frac{\Delta G_m^{i-j}}{k_B T}\right)
\end{align*}
where $\nu_0^i$ is the attempt frequency and $\Delta G_m^{i-j}$ is the free energy of migration encountered by a vacancy jumping from a site of type $i$ to a site of type $j$. The attempt frequency is for all jumps approximated by the frequency of the lowest optical mode at $\Gamma$, $\nu_0\approx 2\,\THz$. \cite{GhoCocWag99}

The probability $\Phi_{ij}$ for a vacancy to jump from a site of type $i$ to a site of type $j$ is given by the fraction of sites of type $j$ in the first nearest neighbor shell of sites of type $i$. Using a simple mapping to index different defect configurations, \MBVcA $\rightarrow $ (1), \MBVcB $\rightarrow$ (2), \MBVab $\rightarrow$ (3), \Vc $\rightarrow$ (4), and \Vab $\rightarrow$ (5) [see \fig{fig:migration_paths}(g) for examples] and taking into account the geometry of the lattice (see \fig{fig:migration_paths}) the following probability matrix is obtained
\begin{align}
  \tensor{\Phi}
  &=
  \frac{1}{8}
  \left(\begin{matrix}
    0       & 0       & 4       & 0           & 4           \\
    0       & 0       & 4       & 0           & 4           \\
    1       & 1       & 2       & 2           & 2           \\    
    0       & 0       & 8\alpha & 0           & 8(1-\alpha) \\
    4\alpha & 4\alpha & 8\alpha & 8(1-\alpha) & 8(1-\alpha)
  \end{matrix}\right)
\label{eq:probmatrix},
\end{align}
where $\alpha=6\fM$.  Here $\fM$ is the fraction of $B$ sites which have been replaced by impurity atoms. The recurrence of the factor eight in \eq{eq:probmatrix} results from the number of oxygen sites in the second neighbor shell of any given oxygen site, while the factor six stems from the number of oxygen sites in the first neighbor shell of a $B$ site. Introducing $\mathrm{W}_{ij}=\Phi_{ij} \mathrm{K}_{ij}$ and $\mathrm{V}_{ij}=\delta_{ij}\sum_k\mathrm{W}_{ik}$, \eq{eq:crate} can be rewritten in a convenient matrix form
\begin{align}
  \vec{\dot{c}}
  &= \left(\tp{\tensor{W}}-\tensor{V}\right) \vec{c},
  \label{eq:model}
\end{align}
which in this work has been numerically
\footnote{
  In principle, the solution of \eq{eq:model} can be written in terms of the eigenvalues and vectors of $\tp{\tensor{W}}-\tensor{V}$. Since the eigenvalues appear in an exponential function, the stability of the solution, which can be tested via \eq{eq:ctot} is highly sensitive to their numerical accuracy. In practice, we have therefore resorted to numerical solvers that approach \eq{eq:model} directly.
}
using an adaptive time step algorithm for stiff differential equations. \footnote{Specifically, we used the \texttt{ode15s} solver of \textsc{matlab}. \cite{matlab}}

It should be noted that the model does not take into account the possibility of two or more oxygen vacancies associating with the same impurity atom. Both experiments and calculations indicate, however, that this is unlikely to occur for the dopants considered in the present work. Similarly the possibility that two or more impurity atoms form an aggregate can be ruled out based on experimental evidence. \cite{MesEicKlo05}

In the present form the model does not include any constraints to allow for the number of oxygen vacancies to be larger than the number of impurity atoms or vice versa. This situation can, however, be implemented rather easily by solving the kinetic model in steps. For instance, consider a case in which the vacancy concentration is $[V_{\O}]=0.01$ and the dopant/impurity concentration is $[M]=0.005$. The sum of the {\em relative} concentrations of complexed vacancies $[M]/[V_\O]$ can, therefore, not exceed $c_{max}=0.5 > c_1 + c_2 + c_3$. Starting from some initial distribution, one solves the kinetic model until $c_1+c_2+c_3$ equals $c_{max}$. At this point all impurities are complexed with vacancies, and one can ``remove'' the free vacancy concentrations $c_4$ and $c_5$ from the model. This is achieved by reducing the $5\times 5$ matrices in Eqs.~(\ref{eq:crate}-\ref{eq:probmatrix}) and \eq{eq:model} to $3\times 3$ matrices, only keeping elements $(i,j)\in \{1,2,3\}$. The opposite scenario, in which the number of impurity/dopant atoms exceeds the number of free vacancies, can be implemented in a similar fashion. For the sake of clarity and because the key conclusions of this work are unaffected by these conditions, we do not consider any of these cases in the remainder part of this paper.

Applying the model to a specific material requires knowledge of the energy differences between various vacancy configurations as well as migration energies. To provide these parameters, we have carried out first-principles calculations that are described in the following section. At this level we neglected the vibrational entropy contribution to the migration barriers and approximated $\Delta G_m \approx \Delta E_m$.

\section{First-principles calculations}
\label{sect:dft}

\subsection{Computational parameters}

The barriers for oxygen vacancy migration in pure as well Cu and Fe-doped lead titanate were calculated within density functional theory (DFT) using the Vienna ab-initio simulation package. \cite{KreHaf93, *KreHaf94, *KreFur96a, *KreFur96b} The potentials due to the ions and the core electrons were represented by the projector-augmented wave method. \cite{Blo94, *KreJou99} The $5d$ electrons of Pb, the $3s$ and $3p$ electrons of Ti as well as the $3p$ electrons of Fe and Cu were treated as part of the valence. The exchange-correlation potential was represented using the local spin density approximation, \cite{CepAld80, *PerZun81}. Supercells containing $2\times 2\times 4$ unit cells equivalent to 80 atoms were employed and the Brillouin zone was sampled using a $2\times 2\times 2$ Monkhorst-Pack mesh. Similar computational parameters were successfully used in previous studies of Cu and Fe-doped lead titanate. \cite{MesEicKlo05, ErhEicTra07, EicErhTra08} For several configurations we also carried out calculations using a $4\times 4\times 4$ mesh and found negligible differences on the order of $0.05\,\eV$ and below. The computations were performed at the theoretical lattice constant of $a_0=3.866\,\text{\AA}$ and the theoretical value for the axial ratio of $c/a=1.05$, both of which are in reasonable agreement with experiment ($a_0=3.905\,\text{\AA}$, $c/a=1.064$ at room temperature, Refs.~\onlinecite{GlaMab78b, RobChe99}). The calculated band gap of 1.47\,eV is considerably smaller than the experimental value, but consistent with the well known band gap error of DFT. As argued in Ref.~\onlinecite{ErhAlb06a} the band gap error is, however, expected to have a minor effect in the context of migration barrier calculations. Migration paths and barriers were determined using the climbing image nudged elastic band method \cite{HenJohJon00, HenUbeJon00} and configurations were optimized until the maximum force was less than 30\,meV/\AA. For charged defects a homogeneous background charge was added.

\subsection{Free oxygen vacancies}
\label{sect:free_VO}

\begin{figure}
  \centering
\includegraphics[width=0.92\linewidth]{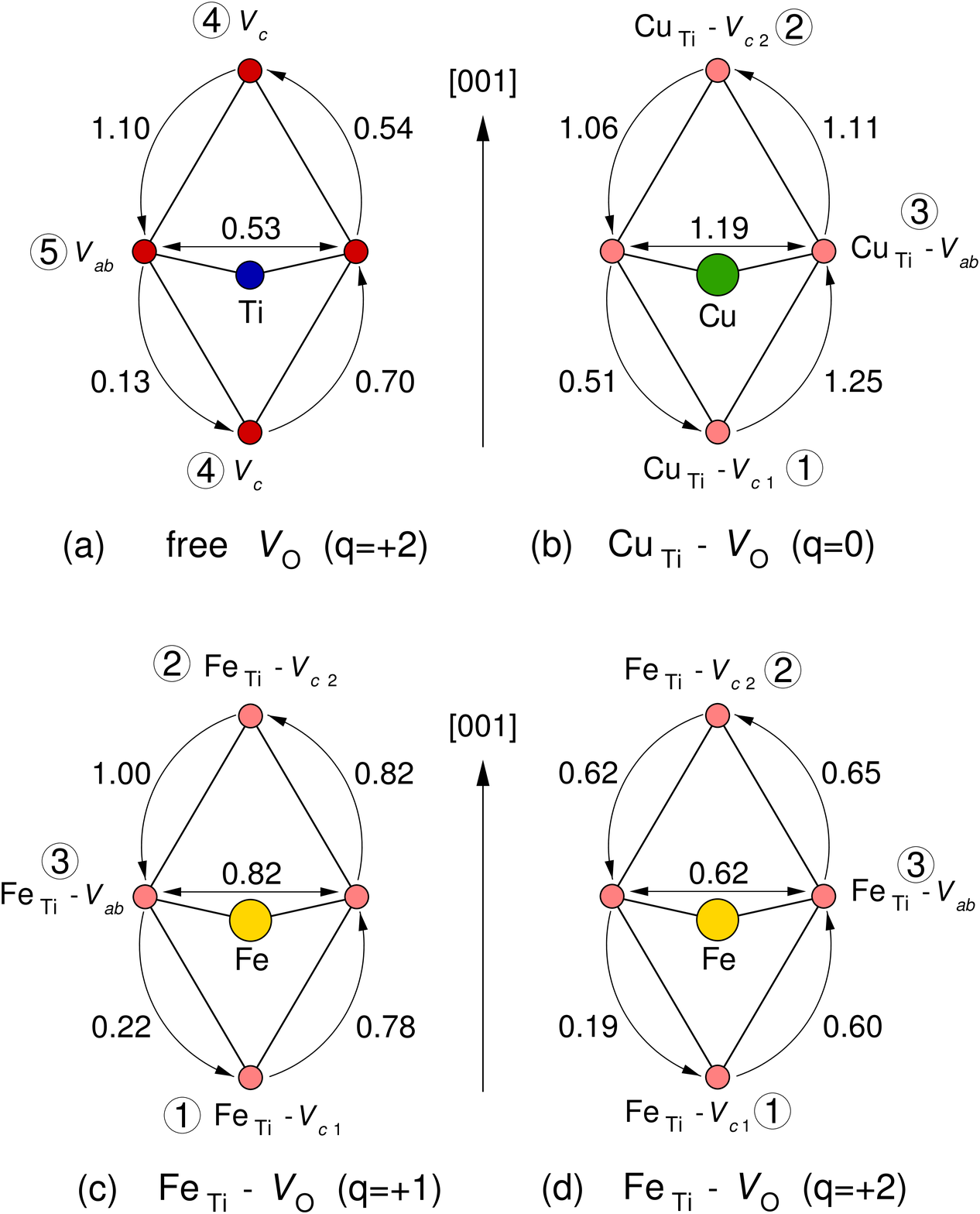}
  \caption{
    Schematic representation of the barriers for the migration of free (a) and complexed (b-d) oxygen vacancies in units of eV. Each figure shows the projection of a $B$O$_6$ octahedron onto the (100) plane. The numbers in circles indicate the indices used to distinguish the different processes. Oxygen sites (and thus possible vacancy sites) are shown as red circles while the position of the $B$-site cation (Ti, Cu, or Fe), which is situated at the center of the oxygen octahedron, is indicated by blue (Ti), green (Cu) and yellow (Fe) circles.
  }
  \label{fig:barriers}
\end{figure}

As indicated in \fig{fig:confs} there are two crystallographically distinct oxygen vacancy sites in the tetragonal perovskite lattice (space group P$4mm$). Vacancies can be situated along the $c$-axis ($\Vc$, Wyckoff site $1b$) or within the $ab$-plane ($\Vab$, Wyckoff site $2c$). \cite{ParCha98, ErhEicTra07}
\footnote{
  Park and Chady \cite{ParCha98} discuss two different configurations for the $ab$-site vacancy, $V_{ab}^{sw}$ and $V_{ab}^{ud}$, but the former one seems to be always lower in energy and therefore prevails. We are thus left with only two different types vacancies, $V_c$ and $V_{ab}$. Also compare the discussion in Ref.~\onlinecite{ErhEicTra07}.
}
Thus, three different migration paths are possible between nearest neighbor sites: pure in-plane migration ($\Vab\rightarrow\Vab$), out-of-plane migration along the positive direction of the tetragonal axis  ($\Vab\rightarrow\Vc~[001]$), and out-of-plane migration along the negative direction of the tetragonal axis ($\Vab\rightarrow\Vc~[00\bar{1}]$).

\begin{table}[bhpt!]
  \caption{
    Migration energies in eV for unbound oxygen vacancies in tetragonal lead titanate. In cases for which the jump is asymmetric both the backward barrier is given in brackets.
  }
  \label{tab:migration_VO}
  \centering
  \begin{tabular}{l*{3}{ll}}
    \hline\hline
    \multicolumn{1}{p{1in}}{Transition}
    & \multicolumn{2}{p{0.75in}}{0}
    & \multicolumn{2}{p{0.75in}}{+1}
    & +2 \\
    \hline
    $V_{ab}\rightarrow V_{ab}$
    & 0.98 &        & 0.62 &        & 0.53 &        \\[6pt]
    $V_{c}\rightarrow V_{ab}~[00\bar{1}]$
    & 0.83 & (0.91) & 0.94 & (0.58) & 1.10 & (0.54) \\[6pt]
    $V_{c}\rightarrow V_{ab}~[001]$
    & 0.50 & (0.59) & 0.58 & (0.22) & 0.70 & (0.13) \\
    \hline\hline
  \end{tabular}
\end{table}

The calculated barriers are compiled in \fig{fig:barriers}(a) and \tab{tab:migration_VO}. In general, the barriers found to be charge state dependent, which is in line with calculations for cubic PbTiO$_3$ \cite{Par03} and cubic BaTiO$_3$. \cite{ErhAlb07} For migration within the $ab$-plane the barriers decrease as electrons are removed from the defect, which is in accord with the calculations on cubic perovskite structures.\cite{Par03, ErhAlb07} The barriers for migration via $c$-type vacancies, in constrast, increase. Since the charge state $q=+2$ prevails vor the oxygen vacancy almost over the entire band gap, \cite{ErhEicTra07} its respective barriers were used in the construction of the energy landscapes used in the kinetic model.

As detailed in the appendix, we can obtain the oxygen diffusivity (excluding the formation energy contribution) from our calculated barriers and compare it with experimental data. We find that the calculated activation energy of 0.7\,eV is in good agreement with the experimental value of 0.87\,eV. \cite{GotHahFle08}

\subsection{Complexed oxygen vacancies}
\label{sect:complexed_VO}

The incorporation of an impurity on the $B$ site breaks  translational symmetry along the tetragonal axis and lifts the degeneracy of the oxygen sites in this direction. One therefore obtains three distinct types of first-neighbor impurity atom--oxygen vacancy associates (compare Figs.~\ref{fig:migration_paths} and \ref{fig:barriers}), which leads to three distinct migration paths. Migration within the second neighbor shell of impurity atoms was not considered, since it has been previously shown that the binding energy between oxygen vacancies and Cu and Fe impurities is the largest in the first neighbor shell. \cite{ErhEicTra07} Thus, once an oxygen vacancy arrives in the vicinity of an impurity via diffusion, it will be attracted to the impurity and eventually reside in its first neighbor shell. The barriers for different paths for the migration of oxygen vacancies in the first nearest neighbor shell of copper and iron impurities are shown in \fig{fig:barriers}(b-d).

\subsection{Construction of energy landscape}
\label{sect:landscape}

\begin{figure*}
  \centering
\includegraphics[width=0.82\linewidth]{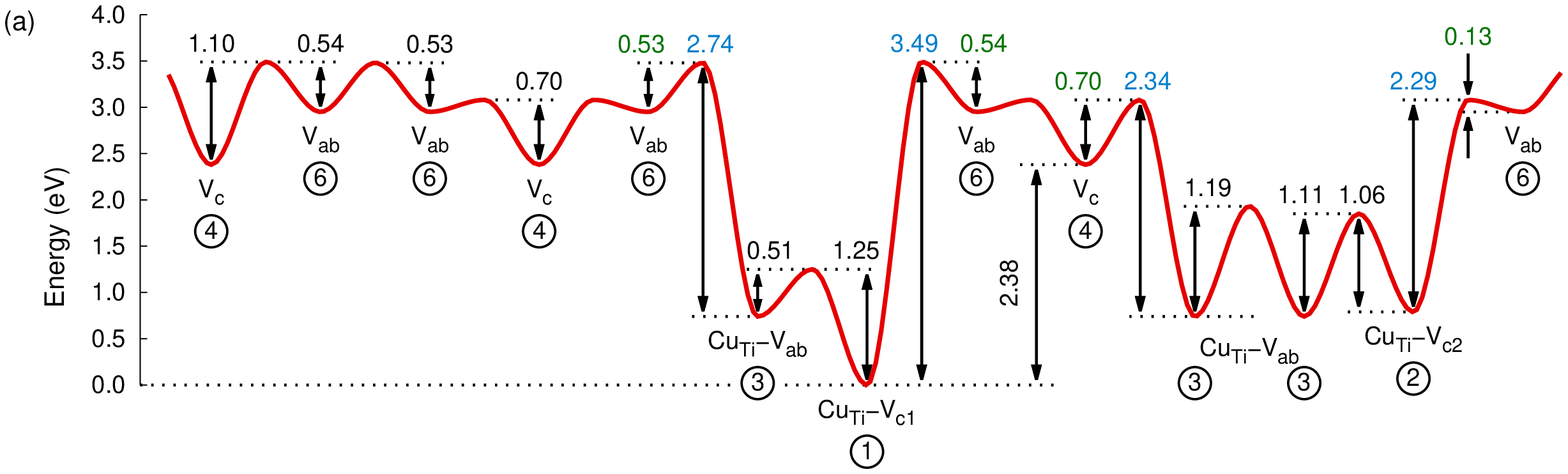}
\includegraphics[width=0.82\linewidth]{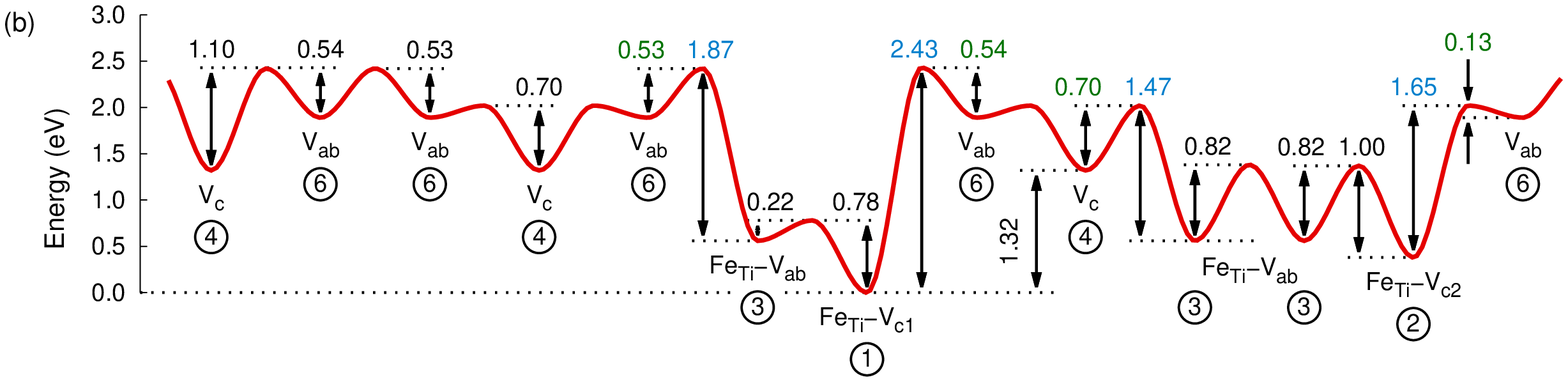}
  \caption{
    Energy surface for the migration of oxygen vacancies in (a) Cu-doped ($\Cu_{\Ti}-V_{\O}$, $q=0$) and (b) Fe-doped PbTiO$_3$ ($\Fe_{\Ti}-V_{\O}$, $q=+1$). The sequence of barriers corresponds to a (hypothetical) continuous trajectory which illustrates all possible migration barriers (compare Figs.~\ref{fig:migration_paths} and \ref{fig:barriers}).
  }
  \label{fig:energy_surface_Cu}
\end{figure*}

So far, we have calculated the migration barriers for free oxygen vacancies, which determine the elements of the rate matrix $\mathrm{K}_{ij}$ for $(i,j)=\{4,5\}$, and the migration barriers for oxygen vacancies in the first neighbor shell of an impurity atoms, which provide the elements with $(i,j)=\{1,2,3\}$. Using these data some parts of the energy landscape can already be constructed as indicated by the migration barriers shown in black in \fig{fig:energy_surface_Cu}.

To determine the barriers for the remaining combinations of $i$ and $j$, e.g., $1-5$ or $3-4$, one would require noticeably larger supercells than the ones employed in the present work. This results from the long ranged Coulombic attraction between the impurity atom and the oxygen vacancy which leads to a gradual transition from a free oxygen vacancy to a vacancy in the first neighbor shell of an impurity ion over several lattice spacings. This complexity can in principle be captured by increasing the dimensionality of the probability and rate matrices, $\Phi$ and $\mathrm{K}$. In the present more simple description, we, however, consider already oxygen vacancies in the second impurity neighbor shell as ``free''.

For completing the rate matrix, we then assume the migration barriers for jumps {\em between free} oxygen vacancies to hold for jumps {\em to complexed} vacancies as well (values marked in green in \fig{fig:energy_surface_Cu}). To determine the barriers for the reverse jumps, we resort to the binding energies calculated for $\Cu_{\Ti}-V_{\O}$ and $\Fe_{\Ti}-V_{\O}$ complexes calculated in Ref.~\onlinecite{ErhEicTra07} (values marked in blue in \fig{fig:energy_surface_Cu}). For Cu and Fe the binding energy amounts to $-2.38\,\eV$ and $-1.32\,\eV$ for Fermi levels near mid gap, which leads to the energy surfaces shown in \fig{fig:energy_surface_Cu}. We have tested the sensitivity of the results of the kinetic model to the barriers for jumps between free and complexed vacancies, which showed the assumptions made in determining their values to be of little consequence.

\section{Results}
\label{sect:results}

In this section we present results obtained using the kinetic model for Cu and Fe-doped lead titanate described in \sect{sect:kinmodel}. To illustrate the general features of vacancy redistribution, we first discuss the results for Cu-doped PbTiO$_3$ in the absence and presence of electric fields in Sects.~\ref{sect:results_dc} and \ref{sect:results_ac}, respectively. The results for Fe-doped material are qualitatively very similar and will be summarized in \sect{sect:Fe_results}.

\subsection{Vacancy redistribution in the absence of electric fields}
\label{sect:results_dc}

\begin{figure}
  \centering
\includegraphics[width=0.85\linewidth]{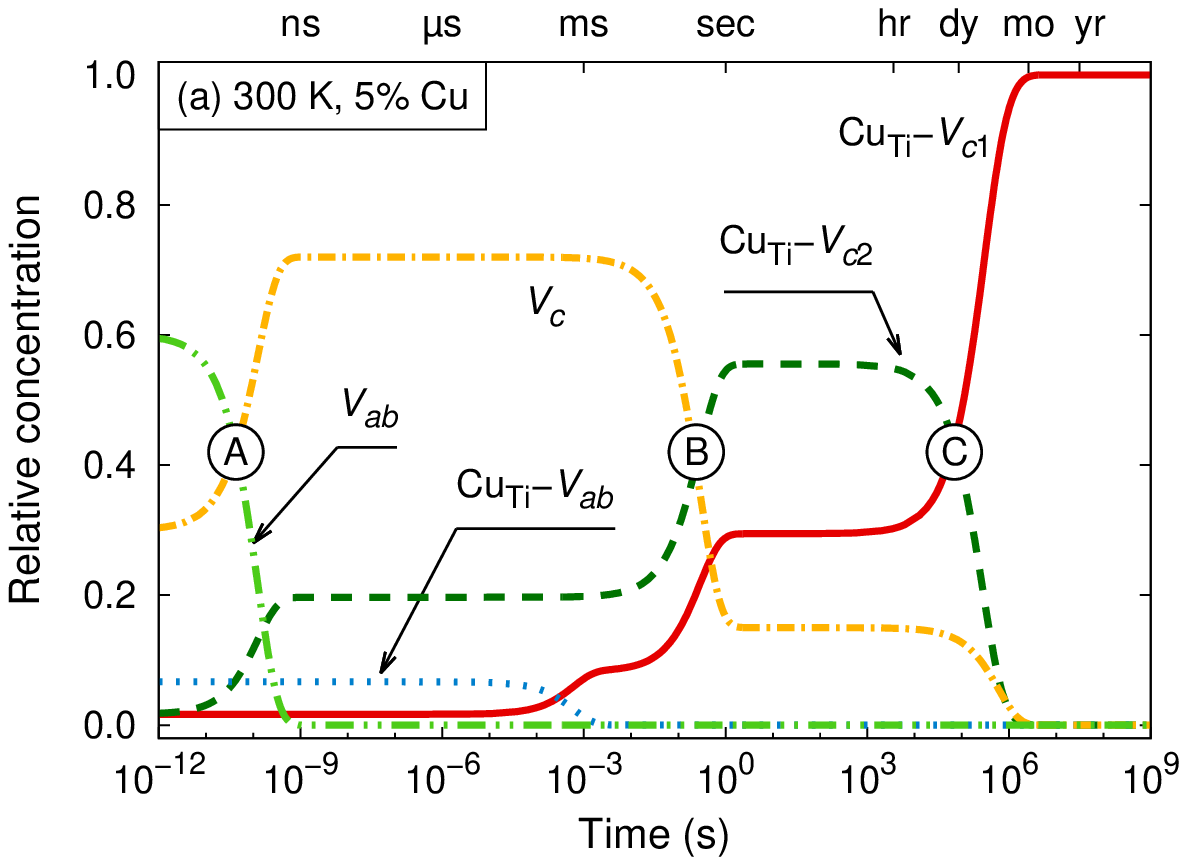}
\includegraphics[width=0.85\linewidth]{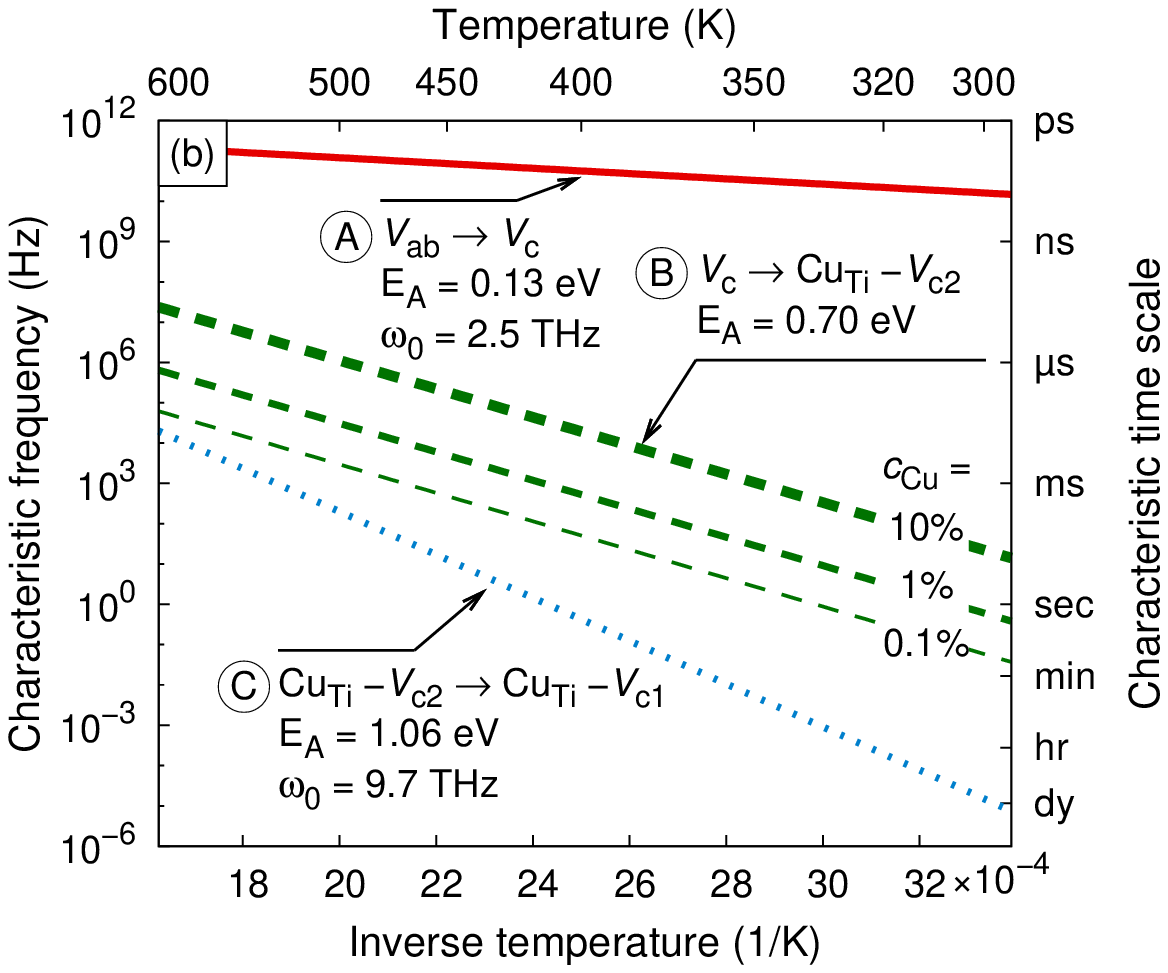}
  \caption{
    Vacancy redistribution in the absence of electric fields for Cu-doped lead titanate:
    (a) Temporal evolution of the relative concentrations of 
different vacancy types at a temperature 300\,K and for a Cu concentration of 5\%. 
    (b) Temperature dependence of the characteristic transition times marked by white circles in (a). The transition time for the conversion from $V_c$ to $\Cu_\Ti-V_{c2}$ depends on the Cu concentration as exemplified by the three green dashed lines of varying thickness.
  }
  \label{fig:Cu_dc}
  \label{fig:chartimes}
\end{figure}

Using the energy landscape for Cu-doped PbTiO$_3$ shown in \fig{fig:energy_surface_Cu} as input data for the kinetic model one can obtain the temporal evolution of the relative concentrations of different types of vacancies as exemplarily shown in \fig{fig:Cu_dc}(a) for a temperature of 300\,K and a dopant concentration of $\fM=5\%$. In this example, the vacancies are initially statistically distributed over all available sites. In \fig{fig:Cu_dc}(a) four distinct time regimes with characteristically different dynamic balance can be identified that are separated by the transitions marked A, B, and C:

\textit{(i)}
The first regime (up to $t\lesssim 10^{-10}\,\s$ at $300\,\K$) is associated with the redistribution of unbound vacancies. As can be seen from \fig{fig:energy_surface_Cu}, \Vc\ is energetically preferred over \Vab. Figure~\ref{fig:Cu_dc}(a) shows that even for a temperature as low as 300\,K the redistribution between these two types of vacancies, {\it i.e.} the installation of the (partial) equilibrium over the subset of unbound vacancies, takes place within fractions of a second.

\textit{(ii)}
During the second stage ($10^{-10}\,\s - 10^0\,\s$ at $300\,\K$) unbound $c$-type vacancies dominate. Concurrently, dopants begin to capture vacancies. The dynamic equilibrium at this point is such that $\Cu_{\Ti}-V_{c2}$ complexes dominate over $\Cu_{\Ti}-V_{c1}$ and $\Cu_{\Ti}-V_{ab}$ dipoles. The former are energetically less favorable but are much more easily accessible since $\Delta G_m(\Vab\rightarrow\MBVcB)$ is only 0.13\,eV compared to $\Delta G_m(\Vab\rightarrow\MBVab)=0.53\,\eV$, $\Delta G_m(\Vc\rightarrow\MBVab)=0.70\,\eV$, and $\Delta G_m(\Vab\rightarrow\MBVcA)=0.54\,\eV$.

\textit{(iii)}
In the third stage ($10^{0}\,\s - 10^5\,\s$ at $300\,\K$) vacancy--dopant complexes take over with $\Cu_{\Ti}-V_{c2}$ being the dominant defect. The prevalence of $\Cu_{\Ti}-V_{c2}$ is inherited from the second stage which determines the initial concentrations for the third stage.

\textit{(iv)}
Eventually, the system reaches equilibrium ($t\gtrsim 10^5\,\s$ at $300\,\K$), {\it i.e.} virtually all vacancies occupy the lowest energy site.
\footnote{
  At finite temperatures a fraction of vacancies will also occupy higher energy configurations. Their number, however, is very small (and indistinguishable from zero on the scale of \fig{fig:Cu_dc}(a)), since it is determined by the Boltzmann factor $\exp(-\Delta E/k_B T)$ where $\Delta E$ is the energy difference between the ground state configuration, $M_{\Ti}-V_{c1}$, and the configuration in question.
}

We can now discuss the ependence of vacancy migration on temperature and dopant concentration. As indicated by the letters A, B, and C in \fig{fig:Cu_dc}(a), characteristic times can readily identified, at which the majority defect type changes. As shown in \fig{fig:chartimes}(b) the temperature dependence of these characteristic times can be fit to an Arrhenius equation, \mbox{$\tau^{-1}=\omega_0 \exp[-E_A/k_B T]$}, using the migration barrier between the states involved as the activation energy. This analysis also demonstrates that the effect of changing the dopant concentration is small and is only visible for the transition between vacancy types $V_c$ and $\Cu_\Ti-V_{c2}$.

Already at temperatures $\gtrsim\!450\,\K$ the full equilibrium is established within less than a second. Since during growth these temperatures are easily reached, the vacancy distribution in tetragonal lead titanate should be in thermal equilibrium, {\it i.e.} virtually all dopants are complexed with vacancies in the ground state configuration, in which the defect dipoles are aligned with the domain polarization.

\subsection{Vacancy redistribution in the presence of electric fields}
\label{sect:results_ac}

In the present model the perturbation introduced by an external electric field enters in two ways. First the barriers for vacancy jumps with components along the direction of the electric field are distorted (``direct effect''), $\Delta G_m\rightarrow \Delta G_m-\delta E$ where $\delta E=E_{loc} \Delta r_{[001]} q e$. Here, $\Delta r_{[001]}$ denotes the displacement along the direction of the electric field which is positive (negative) if the displacement is parallel (anti-parallel) to the electric field; $q$ is the charge state of the defect, $E_{loc}$ is the local electric field, and $e$ denotes the unit charge. Typical electric fields used for poling ferroelectric ceramics are on the order of 2--5\,kV/cm; the local electric field can, however, be larger than this value due to inhomogeneities. \cite{GlaGenKun12} Assuming a value of $E_{loc}=100\,\text{kV/cm}$ and choosing $\Delta r_{[001]}=1\,\text{\AA}$, we obtain an upper limit for $\delta E$ of 0.01\,eV.

Obviously the ``direct'' effect of the electric field is rather small compared to the energy difference between different vacancy types and pertains to charged vacancies only. This implies that with regard to vacancy redistribution a material in a constant external field will behave almost identical to the situation without electric fields.

The situation does, however, change if we consider an oscillating external field. The field induced reversal of the polarization has a much larger effect on the energetics of the system than the direct contribution since reorientation of the polarization implies that the (average) displacement of $B$-site atoms is reversed, thus transforming \MBVcA\ into \MBVcB\ complexes. In the present model, this is equivalent to exchanging rows 1 and 2 of the migration rate matrix, $\mathrm{K}_{ij}$. One can therefore include the effect of an oscillating electric field by (\textit{i}) periodically modifying the barriers for out-of-plane jumps by $\delta E$ and (\textit{ii}) simultaneously exchanging the barriers for jumps involving \MBVcA\ or \MBVcB.

\begin{figure}
  \centering
\includegraphics[width=0.85\linewidth]{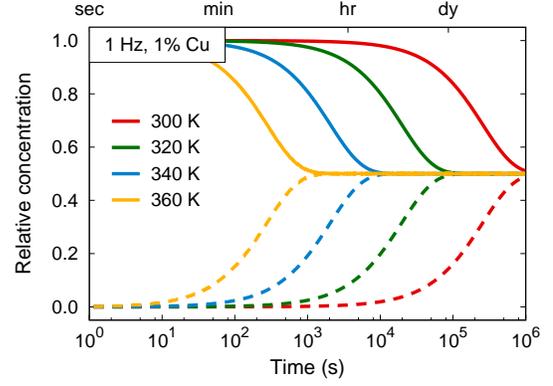}
  \caption{
    Vacancy redistribution in the presence of an oscillating external field:
    Equilibration over vacancy types \MBVcA\ (solid lines) and \MBVcB\ (dashed lines) in the presence of an external oscillating field with  a cycling frequency of 1\,Hz for temperatures between 300 and 360\,K.
  }
  \label{fig:efield_equilibration}
\end{figure}

Figure~\ref{fig:efield_equilibration} illustrates the temporal evolution of the concentrations of \MBVcA\ and \MBVcB\ vacancies in the presence of an oscillating electric field. The plot shows that under prolonged cycling a dynamic balance between the two configurations \MBVcA\ and \MBVcB\ is established. The time after which this balance is obtained depends sensitively on temperature e.g., at room temperature it is reached after about $10^6\,\s$ (approximately two weeks) while at 340\,K it requires only about one hour.


The dynamic balance between \MBVcA\ and \MBVcB\ occurs because the characteristic time required to reach full equilibrium in the absence of external electric fields [see \fig{fig:chartimes}(b)] exceeds the cycling period. In a fast switching field the restoring force of the spontaneous polarization is changing signs on a short time scale. As a result, a mean-field composed of parallel and anti-parallel polarization of the matrix is acting on the dipoles, which evantually populate both orientations along the c-axis. This dynamic equilibrium is sensitive to temperature and the assumption of a static distribution of defect dipoles in a fast switching field should be taken with care. The results show that by applying a bipolar electric field defect dipoles are redistributed and on average the clamping effect on domain walls is reduced. This is in line with recent results on deaging of doped PZT.\cite{GraSuvKun06, GlaGenKun12}

\subsection{Results for Fe-doped lead titanate}
\label{sect:Fe_results}

\begin{figure}
  \centering
\includegraphics[width=0.85\linewidth]{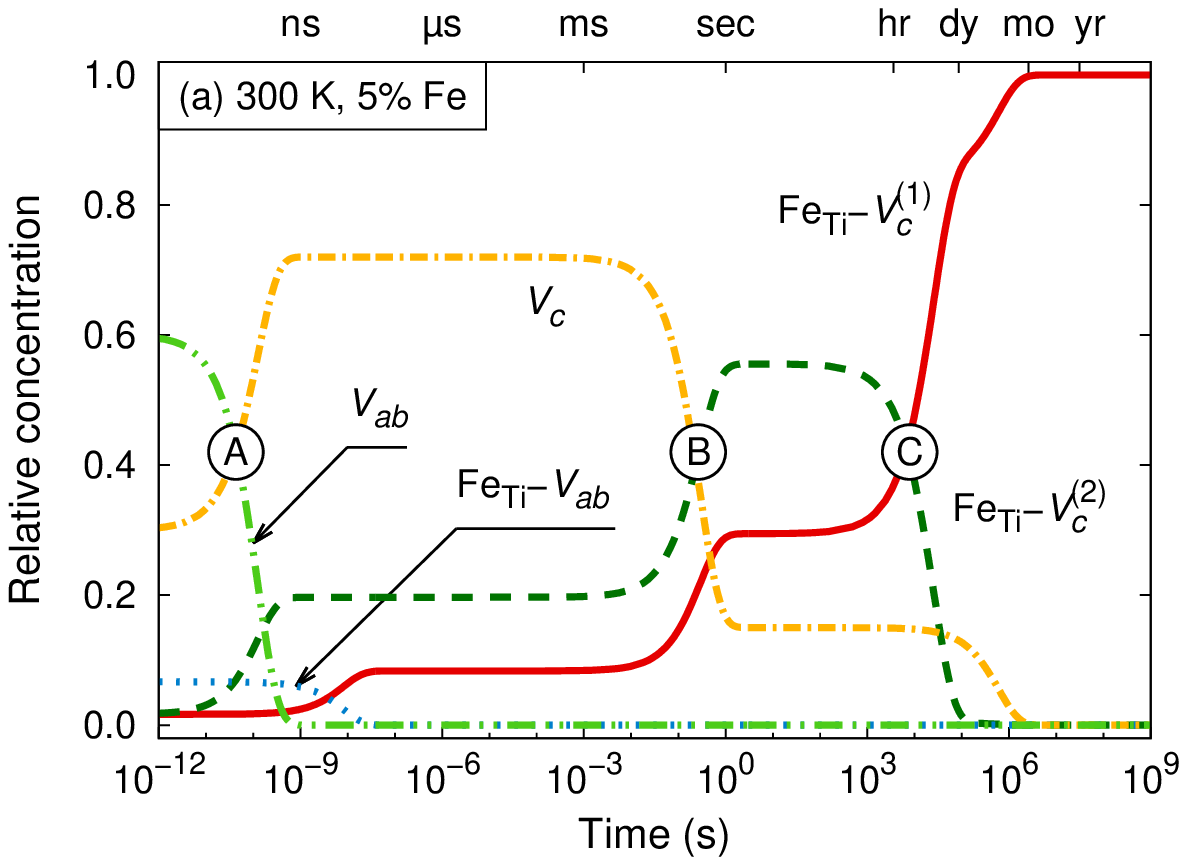}
\includegraphics[width=0.85\linewidth]{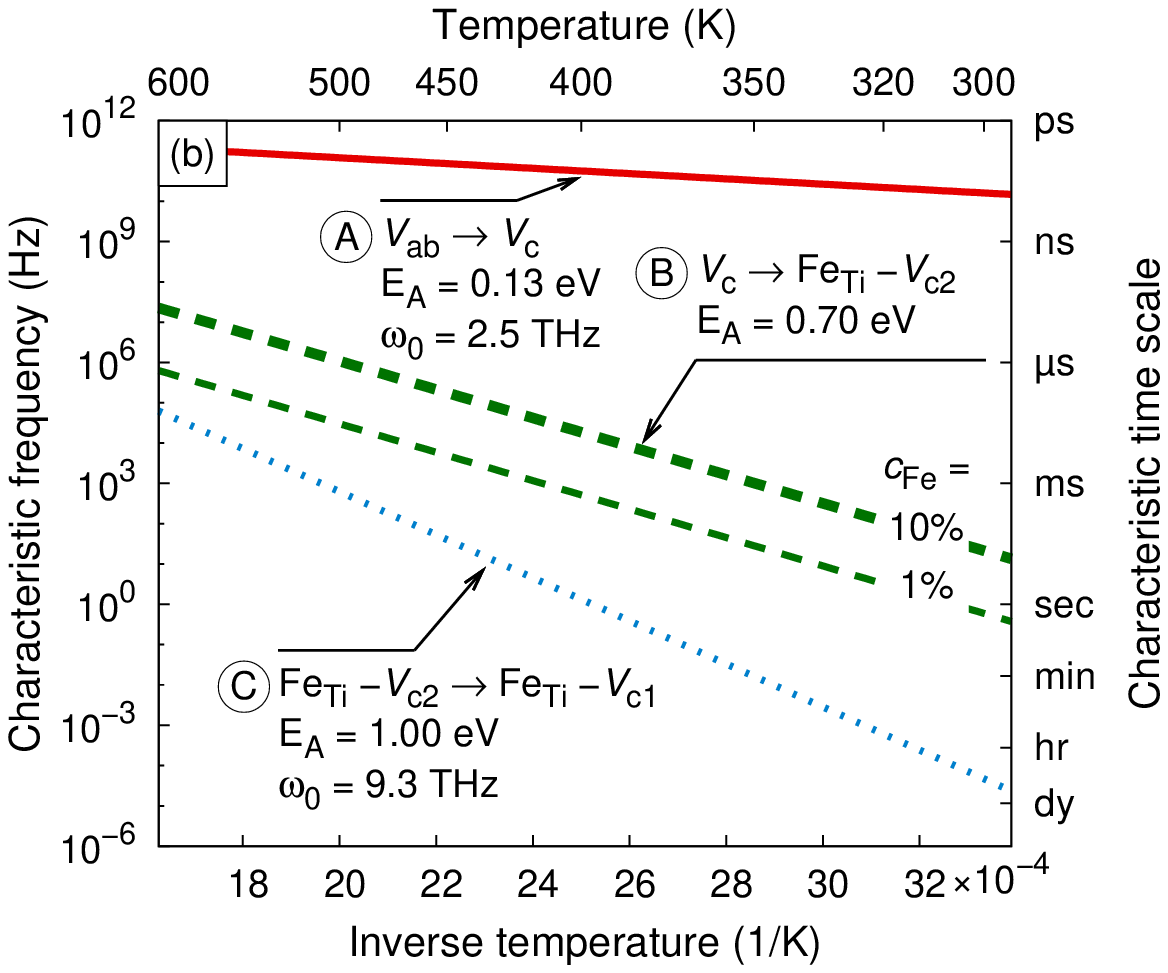}
  \caption{
    (a) Temporal evolution of the relative concentrations of different vacancy types for three different temperatures in Fe-doped lead titante.
    (b) Dependence of characteristic time scales for the transitions indicated by white circles in (a) on temperature and dopant concentration.
  }
  \label{fig:Fe_dc}
  \label{fig:chartimes_Fe}
\end{figure}

Complexes of Fe with oxygen vacancies act as donors leading to electron chemical potentials in the upper half of the band gap. The most stable charge state is $q=+1$. \cite{ErhEicTra07} Under such conditions the binding energy is $-1.32\,\eV$ from which one can construct an energy surface shown in \fig{fig:energy_surface_Cu}(b).

The temporal evolution of different vacancy configurations in the absence of an external electric field is shown in \fig{fig:Fe_dc}(a) which allows us to infer the temperature dependence of the characteristic time scales summarized in \fig{fig:chartimes_Fe}(b). Comparing \fig{fig:Cu_dc} and \fig{fig:Fe_dc} we find that the results for Fe and Cu-doped lead titanate are very similar. This is expected since the first two transitions ($V_{ab}\rightarrow V_{c}$ and $V_c\rightarrow \Fe_\Ti-V_{c2}$, compare \sect{sect:landscape}) are determined by the migration barriers in the pure host. With regard to the third transition between $M_\Ti-V_{c2}$ and $M_\Ti-V_{c1}$ the situation is different as the effective barrier in Fe-doped material is 1.00\,eV and thus slightly smaller than in Cu-doped lead titanate (1.06\,eV, compare \fig{fig:barriers}), which speeds up the transition. We can thus expect that in the presence of an oscillating external field the dynamic equilibrium between $\Fe_\Ti-V_{c2}$ and $\Fe_\Ti-V_{c1}$ is established faster as well, which is confirmed by explicit calculation. Whereas for Cu-doped lead titanate our model calculations predict the equilibrium to be installed over two weeks at room temperature, in Fe-doped material the same process should occur on the order of a day. Similarly at 340\,K equilibration should take only on the order of tens of minutes.

\section{Summary and Discussion}
\label{sect:discussion}

We have parametrized a kinetic model for defect dipole formation and switching by taking data from first-principles calculations for Fe and Cu-doped $\rm PbTiO_3$. We find that  at temperatures $\gtrsim\!450\,\K$, which is well below the Curie temperature of 720\,K, \cite{NohCerIgl95} the formation and alignment of defect dipoles in doped PbTiO$_3$ should occur within less than a second.
\footnote{
  The Curie temperatures of PbTiO$_3$ and PbZrO$_3$ are 720 and 460\,K, respectively, \cite{NohCerIgl95} and the concentration dependent Curie temperature of PZT is bounded by these values.
}

\begin{figure}
  \centering
\includegraphics[width=0.99\columnwidth]{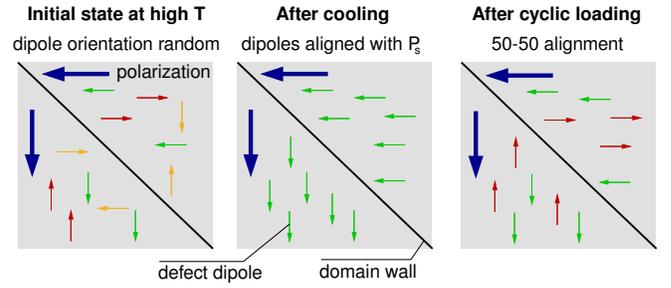}
  \caption{
    Schematic representation of defect dipole arrangements under different conditions as deduced from the kinetic model. The large value arrows indicates the matrix polarization in different domains while the small arrows represent defect dipoles. The thick solid line illustrates the position of a 90\deg\ domain wall.
  }
  \label{fig:compound}
\end{figure}

Bipolar poling leads to a dynamic equilibrium between defect dipoles that are aligned parallel and anti-parallel to the lattice polarization, respectively, and thus can be seen as one major contribution to deaging of PZT ceramics. This is in accord with the experimental observation of deaging by the application of AC fields. \cite{GraSuvKun06, GlaGenKun12} In Cu-doped lead titanate the dynamic equilibration takes about two weeks at 300 K, but is  massively accelerated if temperature is slightly increased. This points to the importance of closely monitoring the sample temperature during testing and studying aging and deaging processes.

For the \MBVcA\ complex, which for both Cu and Fe is the ground state configuration, the local polarization is parallel to the polarization of the surrounding matrix. \cite{ErhEicTra07} Since in contrast \MBVcB\ defects are aligned anti-parallel to the lattice polarization, an increase in their concentration causes an overall loss of switchable polarization. This direct contribution should scale linearly with the number of impurity atoms in the sample but due to the small magnitude of the defect dipole moment will amount to a rather small contribution on the macroscopic scale. Defect dipoles, however, also interact with domain walls and can affect their mobility. In lead titanate and tetragonal PZT one typically observes 90\deg\ domain wall configurations, which is schematically indicated in Fig. \ref{fig:compound}. It has been shown by first-principles calculations that the head-to-tail domain wall configuration shown e.g., in \fig{fig:compound}, is energetically more stable than head-to-head or tail-to-tail configurations. \cite{MeyVan02} In the pristine material after cooling (middle panel of \fig{fig:compound}) all defect dipoles are aligned with the lattice polarization and thus follow the head-to-tail pattern. This situation changes significantly after cyclic loading (right panel of \fig{fig:compound}) since now half of the defect dipoles oppose the lattice polarization and thus create local high-energy head-to-head and tail-to-tail configurations.

Recent simulations of domain wall motion using an empirical force field \cite{ShiGriChe07} have provided impressive evidence that domain wall motion proceeds via a nucleation-and-growth process. It will be subject of future work to determine the role of defect dipoles quantitatively but already at the present stage one can imagine that defect dipoles that are aligned anti-parallel to the lattice polarization in the growing domain will locally impede both nucleation and growth while the opposite can be said for defect dipoles that are aligned parallel to the lattice polarization. One can expect that even though the direct contribution of defect dipoles to the macroscopic polarization is small they can have a significant indirect impact by pinning domain walls and reducing their mobility. It should be stressed that the fact that domain motion occurs via nucleation and growth is crucial in this context since it implies that domain wall motion occurs locally and can thus be strongly influenced by localized defect dipoles.

\section{Conclusions}
\label{sect:conclusions}

In the present work we have derived a kinetic model that allows us to study the temporal evolution of the concentration of different types of vacancies both in the absence and presence of electric fields. The most important input parameter is the energy landscape for oxygen vacancy migration. Using parameters for Cu and Fe-doped \PT\ obtained from density functional theory calculations, we found that the equilibration of the vacancy distribution occurs readily at temperatures considerably below the Curie temperature. As a result in the as-synthesized material virtually all impurity atoms are associated with vacancies forming \MBVcA\ complexes. The complete realignment of vacancy-metal impurity dipoles parallel to the spontaneous polarization  occurs on time scales of hours to days at room temperature, but is massively accelerated if temperature is slightly increased. This provides evidence for the fact that aging due to defect dipoles occurs instantaneously in PbTiO$_3$-based ferroelectrics.

In the presence of an oscillating electric field a dynamic balance between \MBVcA\ and \MBVcB\ is established. Prolonged cycling therefore leads to the accumulation of defect dipoles that oppose the polarization of the encompassing domain. While these defect dipoles directly reduce the switchable polarization, more importantly they can impede domain wall motion, which has been recently shown to proceed via nucleation and growth. \cite{ShiGriChe07}

The present results provide valuable insights into the switching kinetics of defect dipoles in ferroelectrics, which is relevant for understanding aging and deaging mechanisms. The rate equation approach can be adapated straightforwardly to describe more complex geometries and systems. This could be used to model the lattice geometries of e.g., BiFeO$_3$ or LaMnO$_3$.

Kinetic models similar to the one discussed in the present paper can also be used to interpret experimental measurements. To this end, one could employ probes which are sensitive to the orientation of the defect dipoles (e.g., electron spin resonance \cite{MesEicKlo05, EicErhTra08}) and measure the intensity of the signal before, during and after cycling or heat treatments.

\begin{acknowledgments}
This project was partially funded by the \textit{Sonderforschungsbereich 595} ``Fatigue in functional materials'' of the \textit{Deutsche Forschungsgemeinschaft}. P.E. acknowledges funding from the ``Areas of Advance -- Materials Science'' at Chalmers. Com\-puter time allocations by the Swedish National Infrastructure for Computing are gratefully acknowledged.
\end{acknowledgments}

\appendix*

\section{Oxygen Diffusivity}
\label{sect:diffusivity}

We can employ the calculated migration barriers to derive the diffusivity of oxygen vacancies (see e.g., Ref.~\onlinecite{ErhAlb06a}). The rate at which a vacancy jumps along a given path $i$ is
\begin{align}
  \nu &= \nu_0 \exp\left(-\beta \Delta G_i\right)
\end{align}
where $\nu_0$ is the attempt frequency, $\Delta G_i$ is the barrier which has to be surpassed along path $i$, and $\beta=1/k_BT$. Summing over all paths and including the jump lengths $\lambda_i$ as well as the path multiplicities $\zeta_i$ the defect diffusivity is then given by
\begin{align}
  D_d &= \frac{1}{2} \sum_i \nu_i \lambda_i^2 \zeta_i.
  \label{eq:diffusivity}
\end{align}
There are four symmetrically equivalent migration processes ($\zeta=4$) within the $ab$-plane ($\Vab\rightarrow\Vab$), for which the jump lengths projected onto the $ab$-plane and the $c$-axis are $\lambda_{\perp}=a_0$ and $\lambda_{\parallel}=0$, respectively. With regard to of out-of-plane migration ($\Vab\rightarrow\Vc$) there are again four possibilities each for jumps with components along $[001]$ and $[00\bar{1}]$, respectively, associated with displacements $\lambda_{\perp}=a_0/\sqrt{2}$ and $\lambda_{\parallel}=\pm \frac{1}{2}c_0$. Inserting these parameters into \eq{eq:diffusivity} yields
\begin{align*}
  D_{\perp}
  &= a_0^2 \nu_0
  \bigg[
  \exp\left(-\beta\Delta G_{ab-c}^{[001]}\right)
  + \exp\left(-\beta\Delta G_{ab-c}^{[00\bar{1}]}\right)
  \\
  &\quad\quad\qquad
  + 2 \exp\left(-\beta\Delta G_{ab-ab}\right)
  \exp\left(-\beta\Delta G_{ab-c}^{form}\right)
  \bigg]
  \\
  D_{\parallel}
  &= \frac{1}{2} c_0^2 \nu_0
  \left[
  \exp\left(-\beta\Delta G_{ab-c}^{[001]}\right)
  + \exp\left(-\beta\Delta G_{ab-c}^{[00\bar{1}]}\right)
  \right]
\end{align*}
where the very last term takes into account the equilibrium occupancy of $ab$-sites with respect to $c$-site vacancies and $\Delta G_{ab-c}^{form}$ is the difference between the formation free energies of $c$ and $ab$-site vacancies. Finally, the isotropic diffusivity is given as the trace of the diffusivity tensor which for tetragonal symmetry yields $D = 2 D_{\perp} + D_{\parallel}$.

According to \tab{tab:migration_VO} the process $\Vc\rightarrow\Vab [001]$ has the lowest barrier for all charge states (also compare the barriers between the four leftmost minima in \fig{fig:energy_surface_Cu}). Therefore, the isotropic defect diffusivity is approximately
\begin{align}
  D &\approx
  \left( 2 a_0^2 + \frac{1}{2} c_0^2 \right)
  \nu_0 \exp\left(-\beta\Delta G_{ab-c}^{[001]}\right).
\end{align}
Using the migration barriers for charge state $q=+2$, approximating the attempt frequency by the lowest optical mode  at $\Gamma$, $\nu_0\approx 2\,\THz$, \cite{GhoCocWag99} and using the experimental lattice constants one obtains
\begin{align*}
  D &\approx 7.8 \times 10^{-3}
  \exp\left(-0.7\,\eV/k_B T\right) \cm^2/\s.
\end{align*}
The activation barrier in this expression is in reasonable agreement with recent diffusion measurements on lead titanate-zirconate (PZT) alloys \cite{GotHahFle08}, in which the migration barrier for oxygen vacancy migration was found to be $0.87\,\eV$.

It should be noted that the experiments in Ref.~\onlinecite{GotHahFle08} were carried out at low temperatures on samples that contained an extrinsic concentration of oxygen vacancies. As a result, the calculated and measured pre-factors cannot be directly compared since the latter contains the (unknown) concentration of extrinsic vacancies in the samples.


\begin{thebibliography}{57}%
\makeatletter
\providecommand \@ifxundefined [1]{%
 \@ifx{#1\undefined}
}%
\providecommand \@ifnum [1]{%
 \ifnum #1\expandafter \@firstoftwo
 \else \expandafter \@secondoftwo
 \fi
}%
\providecommand \@ifx [1]{%
 \ifx #1\expandafter \@firstoftwo
 \else \expandafter \@secondoftwo
 \fi
}%
\providecommand \natexlab [1]{#1}%
\providecommand \enquote  [1]{``#1''}%
\providecommand \bibnamefont  [1]{#1}%
\providecommand \bibfnamefont [1]{#1}%
\providecommand \citenamefont [1]{#1}%
\providecommand \href@noop [0]{\@secondoftwo}%
\providecommand \href [0]{\begingroup \@sanitize@url \@href}%
\providecommand \@href[1]{\@@startlink{#1}\@@href}%
\providecommand \@@href[1]{\endgroup#1\@@endlink}%
\providecommand \@sanitize@url [0]{\catcode `\\12\catcode `\$12\catcode
  `\&12\catcode `\#12\catcode `\^12\catcode `\_12\catcode `\%12\relax}%
\providecommand \@@startlink[1]{}%
\providecommand \@@endlink[0]{}%
\providecommand \url  [0]{\begingroup\@sanitize@url \@url }%
\providecommand \@url [1]{\endgroup\@href {#1}{\urlprefix }}%
\providecommand \urlprefix  [0]{URL }%
\providecommand \Eprint [0]{\href }%
\providecommand \doibase [0]{http://dx.doi.org/}%
\providecommand \selectlanguage [0]{\@gobble}%
\providecommand \bibinfo  [0]{\@secondoftwo}%
\providecommand \bibfield  [0]{\@secondoftwo}%
\providecommand \translation [1]{[#1]}%
\providecommand \BibitemOpen [0]{}%
\providecommand \bibitemStop [0]{}%
\providecommand \bibitemNoStop [0]{.\EOS\space}%
\providecommand \EOS [0]{\spacefactor3000\relax}%
\providecommand \BibitemShut  [1]{\csname bibitem#1\endcsname}%
\let\auto@bib@innerbib\@empty
\bibitem [{\citenamefont {Plessner}(1956)}]{Ple56}%
  \BibitemOpen
  \bibfield  {author} {\bibinfo {author} {\bibfnamefont {K.~W.}\ \bibnamefont
  {Plessner}},\ }\href {\doibase 10.1088/0370-1301/69/12/309} {\bibfield
  {journal} {\bibinfo  {journal} {Proc. Phys. Soc. B (London)}\ }\textbf
  {\bibinfo {volume} {69}},\ \bibinfo {pages} {1261} (\bibinfo {year}
  {1956})}\BibitemShut {NoStop}%
\bibitem [{\citenamefont {Ikegami}\ and\ \citenamefont
  {Ueda}(1967)}]{IkeUed67}%
  \BibitemOpen
  \bibfield  {author} {\bibinfo {author} {\bibfnamefont {S.}~\bibnamefont
  {Ikegami}}\ and\ \bibinfo {author} {\bibfnamefont {I.}~\bibnamefont {Ueda}},\
  }\href {\doibase 10.1143/JPSJ.22.725} {\bibfield  {journal} {\bibinfo
  {journal} {J. Phys. Soc. Japan}\ }\textbf {\bibinfo {volume} {22}},\ \bibinfo
  {pages} {725} (\bibinfo {year} {1967})}\BibitemShut {NoStop}%
\bibitem [{\citenamefont {Carl}\ and\ \citenamefont
  {H\"ardtl}(1978)}]{CarHar78}%
  \BibitemOpen
  \bibfield  {author} {\bibinfo {author} {\bibfnamefont {K.}~\bibnamefont
  {Carl}}\ and\ \bibinfo {author} {\bibfnamefont {K.}~\bibnamefont
  {H\"ardtl}},\ }\href {\doibase 10.1080/00150197808236770} {\bibfield
  {journal} {\bibinfo  {journal} {Ferroelectrics}\ }\textbf {\bibinfo {volume}
  {17}},\ \bibinfo {pages} {473} (\bibinfo {year} {1978})}\BibitemShut
  {NoStop}%
\bibitem [{\citenamefont {Takahashi}(1982)}]{Tak82}%
  \BibitemOpen
  \bibfield  {author} {\bibinfo {author} {\bibfnamefont {S.}~\bibnamefont
  {Takahashi}},\ }\href {\doibase 10.1080/00150198208210617} {\bibfield
  {journal} {\bibinfo  {journal} {Ferroelectrics}\ }\textbf {\bibinfo {volume}
  {41}},\ \bibinfo {pages} {143} (\bibinfo {year} {1982})}\BibitemShut
  {NoStop}%
\bibitem [{\citenamefont {Arlt}\ and\ \citenamefont
  {Neumann}(1988)}]{ArlNeu88}%
  \BibitemOpen
  \bibfield  {author} {\bibinfo {author} {\bibfnamefont {G.}~\bibnamefont
  {Arlt}}\ and\ \bibinfo {author} {\bibfnamefont {H.}~\bibnamefont {Neumann}},\
  }\href {\doibase 10.1080/00150198808201374} {\bibfield  {journal} {\bibinfo
  {journal} {Ferroelectrics}\ }\textbf {\bibinfo {volume} {87}},\ \bibinfo
  {pages} {109} (\bibinfo {year} {1988})}\BibitemShut {NoStop}%
\bibitem [{\citenamefont {Lohk\"amper}\ \emph {et~al.}(1990)\citenamefont
  {Lohk\"amper}, \citenamefont {Neumann},\ and\ \citenamefont
  {Arlt}}]{LohNeuArl90}%
  \BibitemOpen
  \bibfield  {author} {\bibinfo {author} {\bibfnamefont {R.}~\bibnamefont
  {Lohk\"amper}}, \bibinfo {author} {\bibfnamefont {H.}~\bibnamefont
  {Neumann}}, \ and\ \bibinfo {author} {\bibfnamefont {G.}~\bibnamefont
  {Arlt}},\ }\href {\doibase 10.1063/1.346212} {\bibfield  {journal} {\bibinfo
  {journal} {J. Appl. Phys.}\ }\textbf {\bibinfo {volume} {68}},\ \bibinfo
  {pages} {4220} (\bibinfo {year} {1990})}\BibitemShut {NoStop}%
\bibitem [{\citenamefont {Warren}\ \emph {et~al.}(1995)\citenamefont {Warren},
  \citenamefont {Dimos}, \citenamefont {Pike}, \citenamefont {Vanheusden},\
  and\ \citenamefont {Ramesh}}]{WarDimPik95}%
  \BibitemOpen
  \bibfield  {author} {\bibinfo {author} {\bibfnamefont {W.~L.}\ \bibnamefont
  {Warren}}, \bibinfo {author} {\bibfnamefont {D.}~\bibnamefont {Dimos}},
  \bibinfo {author} {\bibfnamefont {G.~E.}\ \bibnamefont {Pike}}, \bibinfo
  {author} {\bibfnamefont {K.}~\bibnamefont {Vanheusden}}, \ and\ \bibinfo
  {author} {\bibfnamefont {R.}~\bibnamefont {Ramesh}},\ }\href {\doibase
  10.1063/1.115058} {\bibfield  {journal} {\bibinfo  {journal} {Appl. Phys.
  Lett.}\ }\textbf {\bibinfo {volume} {67}},\ \bibinfo {pages} {1689} (\bibinfo
  {year} {1995})}\BibitemShut {NoStop}%
\bibitem [{\citenamefont {Afanasjev}\ \emph {et~al.}(2001)\citenamefont
  {Afanasjev}, \citenamefont {Petrov}, \citenamefont {Pronin}, \citenamefont
  {Tarakanov}, \citenamefont {Kaptelov},\ and\ \citenamefont
  {Graul}}]{AfaPetPro01}%
  \BibitemOpen
  \bibfield  {author} {\bibinfo {author} {\bibfnamefont {V.~P.}\ \bibnamefont
  {Afanasjev}}, \bibinfo {author} {\bibfnamefont {A.~A.}\ \bibnamefont
  {Petrov}}, \bibinfo {author} {\bibfnamefont {I.~P.}\ \bibnamefont {Pronin}},
  \bibinfo {author} {\bibfnamefont {E.~A.}\ \bibnamefont {Tarakanov}}, \bibinfo
  {author} {\bibfnamefont {E.~J.}\ \bibnamefont {Kaptelov}}, \ and\ \bibinfo
  {author} {\bibfnamefont {J.}~\bibnamefont {Graul}},\ }\href {\doibase
  10.1088/0953-8984/13/39/304} {\bibfield  {journal} {\bibinfo  {journal} {J.
  Phys. Cond. Matter}\ }\textbf {\bibinfo {volume} {13}},\ \bibinfo {pages}
  {8755} (\bibinfo {year} {2001})}\BibitemShut {NoStop}%
\bibitem [{\citenamefont {Zhang}\ and\ \citenamefont {Ren}(2005)}]{ZhaRen05}%
  \BibitemOpen
  \bibfield  {author} {\bibinfo {author} {\bibfnamefont {L.~X.}\ \bibnamefont
  {Zhang}}\ and\ \bibinfo {author} {\bibfnamefont {X.}~\bibnamefont {Ren}},\
  }\href {\doibase 10.1103/PhysRevB.71.174108} {\bibfield  {journal} {\bibinfo
  {journal} {Phys. Rev. B}\ }\textbf {\bibinfo {volume} {71}},\ \bibinfo
  {pages} {174108} (\bibinfo {year} {2005})}\BibitemShut {NoStop}%
\bibitem [{\citenamefont {Zhang}\ \emph {et~al.}(2008)\citenamefont {Zhang},
  \citenamefont {Erdem}, \citenamefont {Ren},\ and\ \citenamefont
  {Eichel}}]{ZhaErdRen08}%
  \BibitemOpen
  \bibfield  {author} {\bibinfo {author} {\bibfnamefont {L.}~\bibnamefont
  {Zhang}}, \bibinfo {author} {\bibfnamefont {E.}~\bibnamefont {Erdem}},
  \bibinfo {author} {\bibfnamefont {X.}~\bibnamefont {Ren}}, \ and\ \bibinfo
  {author} {\bibfnamefont {R.-A.}\ \bibnamefont {Eichel}},\ }\href {\doibase
  10.1063/1.3006327} {\bibfield  {journal} {\bibinfo  {journal} {Appl. Phys.
  Lett.}\ }\textbf {\bibinfo {volume} {93}},\ \bibinfo {pages} {202901}
  (\bibinfo {year} {2008})}\BibitemShut {NoStop}%
\bibitem [{\citenamefont {Morozov}\ and\ \citenamefont
  {Damjanovic}(2008)}]{MorDam08}%
  \BibitemOpen
  \bibfield  {author} {\bibinfo {author} {\bibfnamefont {M.~I.}\ \bibnamefont
  {Morozov}}\ and\ \bibinfo {author} {\bibfnamefont {D.}~\bibnamefont
  {Damjanovic}},\ }\href {\doibase 10.1063/1.2963704} {\bibfield  {journal}
  {\bibinfo  {journal} {J. Appl. Phys.}\ }\textbf {\bibinfo {volume} {104}},\
  \bibinfo {pages} {034107} (\bibinfo {year} {2008})}\BibitemShut {NoStop}%
\bibitem [{\citenamefont {Genenko}\ \emph {et~al.}(2009)\citenamefont
  {Genenko}, \citenamefont {Glaum}, \citenamefont {Hirsch}, \citenamefont
  {Kungl}, \citenamefont {Hoffmann},\ and\ \citenamefont
  {Granzow}}]{GenGlaHir09}%
  \BibitemOpen
  \bibfield  {author} {\bibinfo {author} {\bibfnamefont {Y.~A.}\ \bibnamefont
  {Genenko}}, \bibinfo {author} {\bibfnamefont {J.}~\bibnamefont {Glaum}},
  \bibinfo {author} {\bibfnamefont {O.}~\bibnamefont {Hirsch}}, \bibinfo
  {author} {\bibfnamefont {H.}~\bibnamefont {Kungl}}, \bibinfo {author}
  {\bibfnamefont {M.~J.}\ \bibnamefont {Hoffmann}}, \ and\ \bibinfo {author}
  {\bibfnamefont {T.}~\bibnamefont {Granzow}},\ }\href {\doibase
  10.1103/PhysRevB.80.224109} {\bibfield  {journal} {\bibinfo  {journal} {Phys.
  Rev. B}\ }\textbf {\bibinfo {volume} {80}},\ \bibinfo {pages} {224109}
  (\bibinfo {year} {2009})}\BibitemShut {NoStop}%
\bibitem [{\citenamefont {Zhang}\ and\ \citenamefont {Ren}(2010)}]{ZhaRen10}%
  \BibitemOpen
  \bibfield  {author} {\bibinfo {author} {\bibfnamefont {L.}~\bibnamefont
  {Zhang}}\ and\ \bibinfo {author} {\bibfnamefont {X.}~\bibnamefont {Ren}},\
  }\href {\doibase 10.1142/S1793604710000890} {\bibfield  {journal} {\bibinfo
  {journal} {Func. Mater. Lett.}\ }\textbf {\bibinfo {volume} {3}},\ \bibinfo
  {pages} {69} (\bibinfo {year} {2010})}\BibitemShut {NoStop}%
\bibitem [{\citenamefont {Tagantsev}\ \emph {et~al.}(2001)\citenamefont
  {Tagantsev}, \citenamefont {Stolichnov}, \citenamefont {Colla},\ and\
  \citenamefont {Setter}}]{TagStoCol01}%
  \BibitemOpen
  \bibfield  {author} {\bibinfo {author} {\bibfnamefont {A.}~\bibnamefont
  {Tagantsev}}, \bibinfo {author} {\bibfnamefont {I.}~\bibnamefont
  {Stolichnov}}, \bibinfo {author} {\bibfnamefont {E.}~\bibnamefont {Colla}}, \
  and\ \bibinfo {author} {\bibfnamefont {N.}~\bibnamefont {Setter}},\ }\href
  {\doibase 10.1063/1.1381542} {\bibfield  {journal} {\bibinfo  {journal} {J.
  Appl. Phys.}\ }\textbf {\bibinfo {volume} {90}},\ \bibinfo {pages} {1387}
  (\bibinfo {year} {2001})}\BibitemShut {NoStop}%
\bibitem [{\citenamefont {Ren}(2004)}]{Ren04}%
  \BibitemOpen
  \bibfield  {author} {\bibinfo {author} {\bibfnamefont {X.}~\bibnamefont
  {Ren}},\ }\href {\doibase 10.1038/nmat1051} {\bibfield  {journal} {\bibinfo
  {journal} {Nature Mater.}\ }\textbf {\bibinfo {volume} {3}},\ \bibinfo
  {pages} {91} (\bibinfo {year} {2004})}\BibitemShut {NoStop}%
\bibitem [{\citenamefont {Shin}\ \emph {et~al.}(2007)\citenamefont {Shin},
  \citenamefont {Grinberg}, \citenamefont {Chen},\ and\ \citenamefont
  {Rappe}}]{ShiGriChe07}%
  \BibitemOpen
  \bibfield  {author} {\bibinfo {author} {\bibfnamefont {Y.-H.}\ \bibnamefont
  {Shin}}, \bibinfo {author} {\bibfnamefont {I.}~\bibnamefont {Grinberg}},
  \bibinfo {author} {\bibfnamefont {I.-W.}\ \bibnamefont {Chen}}, \ and\
  \bibinfo {author} {\bibfnamefont {A.~M.}\ \bibnamefont {Rappe}},\ }\href
  {\doibase 10.1038/nature06165} {\bibfield  {journal} {\bibinfo  {journal}
  {Nature}\ }\textbf {\bibinfo {volume} {449}},\ \bibinfo {pages} {881}
  (\bibinfo {year} {2007})}\BibitemShut {NoStop}%
\bibitem [{\citenamefont {Jia}\ \emph {et~al.}(2008)\citenamefont {Jia},
  \citenamefont {Mi}, \citenamefont {Urban}, \citenamefont {Vrejoiu},
  \citenamefont {Alexe},\ and\ \citenamefont {Hesse}}]{JiaMiUrb08}%
  \BibitemOpen
  \bibfield  {author} {\bibinfo {author} {\bibfnamefont {C.-L.}\ \bibnamefont
  {Jia}}, \bibinfo {author} {\bibfnamefont {S.-B.}\ \bibnamefont {Mi}},
  \bibinfo {author} {\bibfnamefont {K.}~\bibnamefont {Urban}}, \bibinfo
  {author} {\bibfnamefont {I.}~\bibnamefont {Vrejoiu}}, \bibinfo {author}
  {\bibfnamefont {M.}~\bibnamefont {Alexe}}, \ and\ \bibinfo {author}
  {\bibfnamefont {D.}~\bibnamefont {Hesse}},\ }\href {\doibase
  10.1038/nmat2080} {\bibfield  {journal} {\bibinfo  {journal} {Nature Mater.}\
  }\textbf {\bibinfo {volume} {7}},\ \bibinfo {pages} {57} (\bibinfo {year}
  {2008})}\BibitemShut {NoStop}%
\bibitem [{\citenamefont {Erhart}\ \emph {et~al.}(2007)\citenamefont {Erhart},
  \citenamefont {Eichel}, \citenamefont {Tr\"askelin},\ and\ \citenamefont
  {Albe}}]{ErhEicTra07}%
  \BibitemOpen
  \bibfield  {author} {\bibinfo {author} {\bibfnamefont {P.}~\bibnamefont
  {Erhart}}, \bibinfo {author} {\bibfnamefont {R.-A.}\ \bibnamefont {Eichel}},
  \bibinfo {author} {\bibfnamefont {P.}~\bibnamefont {Tr\"askelin}}, \ and\
  \bibinfo {author} {\bibfnamefont {K.}~\bibnamefont {Albe}},\ }\href {\doibase
  10.1103/PhysRevB.76.174116} {\bibfield  {journal} {\bibinfo  {journal} {Phys.
  Rev. B}\ }\textbf {\bibinfo {volume} {76}},\ \bibinfo {pages} {174116}
  (\bibinfo {year} {2007})}\BibitemShut {NoStop}%
\bibitem [{\citenamefont {P\"oykk\"o}\ and\ \citenamefont
  {Chadi}(1999)}]{PoyCha99a}%
  \BibitemOpen
  \bibfield  {author} {\bibinfo {author} {\bibfnamefont {S.}~\bibnamefont
  {P\"oykk\"o}}\ and\ \bibinfo {author} {\bibfnamefont {D.~J.}\ \bibnamefont
  {Chadi}},\ }\href {\doibase 10.1103/PhysRevLett.83.1231} {\bibfield
  {journal} {\bibinfo  {journal} {Phys. Rev. Lett.}\ }\textbf {\bibinfo
  {volume} {83}},\ \bibinfo {pages} {1231} (\bibinfo {year}
  {1999})}\BibitemShut {NoStop}%
\bibitem [{\citenamefont {Me\v{s}tri\'c}\ \emph {et~al.}(2004)\citenamefont
  {Me\v{s}tri\'c}, \citenamefont {Eichel}, \citenamefont {Dinse}, \citenamefont
  {Ozarowski}, \citenamefont {van Tol},\ and\ \citenamefont
  {Brunel}}]{MesEicDin04}%
  \BibitemOpen
  \bibfield  {author} {\bibinfo {author} {\bibfnamefont {H.}~\bibnamefont
  {Me\v{s}tri\'c}}, \bibinfo {author} {\bibfnamefont {R.-A.}\ \bibnamefont
  {Eichel}}, \bibinfo {author} {\bibfnamefont {K.-P.}\ \bibnamefont {Dinse}},
  \bibinfo {author} {\bibfnamefont {A.}~\bibnamefont {Ozarowski}}, \bibinfo
  {author} {\bibfnamefont {J.}~\bibnamefont {van Tol}}, \ and\ \bibinfo
  {author} {\bibfnamefont {L.}~\bibnamefont {Brunel}},\ }\href {\doibase
  10.1063/1.1808477} {\bibfield  {journal} {\bibinfo  {journal} {J. Appl.
  Phys.}\ }\textbf {\bibinfo {volume} {96}},\ \bibinfo {pages} {7440} (\bibinfo
  {year} {2004})}\BibitemShut {NoStop}%
\bibitem [{\citenamefont {Me\v{s}tri\'{c}}\ \emph {et~al.}(2005)\citenamefont
  {Me\v{s}tri\'{c}}, \citenamefont {Eichel}, \citenamefont {Kloss},
  \citenamefont {Dinse}, \citenamefont {Laubach}, \citenamefont {Laubach},
  \citenamefont {Schmidt}, \citenamefont {Sch\"onau}, \citenamefont {Knapp},\
  and\ \citenamefont {Ehrenberg}}]{MesEicKlo05}%
  \BibitemOpen
  \bibfield  {author} {\bibinfo {author} {\bibfnamefont {H.}~\bibnamefont
  {Me\v{s}tri\'{c}}}, \bibinfo {author} {\bibfnamefont {R.-A.}\ \bibnamefont
  {Eichel}}, \bibinfo {author} {\bibfnamefont {T.}~\bibnamefont {Kloss}},
  \bibinfo {author} {\bibfnamefont {K.-P.}\ \bibnamefont {Dinse}}, \bibinfo
  {author} {\bibfnamefont {S.}~\bibnamefont {Laubach}}, \bibinfo {author}
  {\bibfnamefont {S.}~\bibnamefont {Laubach}}, \bibinfo {author} {\bibfnamefont
  {P.~C.}\ \bibnamefont {Schmidt}}, \bibinfo {author} {\bibfnamefont {K.~A.}\
  \bibnamefont {Sch\"onau}}, \bibinfo {author} {\bibfnamefont {M.}~\bibnamefont
  {Knapp}}, \ and\ \bibinfo {author} {\bibfnamefont {H.}~\bibnamefont
  {Ehrenberg}},\ }\href {\doibase 10.1103/PhysRevB.71.134109} {\bibfield
  {journal} {\bibinfo  {journal} {Phys. Rev. B}\ }\textbf {\bibinfo {volume}
  {71}},\ \bibinfo {pages} {134109} (\bibinfo {year} {2005})}\BibitemShut
  {NoStop}%
\bibitem [{\citenamefont {Boonchun}\ \emph {et~al.}(2007)\citenamefont
  {Boonchun}, \citenamefont {Smith}, \citenamefont {Cherdhirunkorn},\ and\
  \citenamefont {Limpijumnong}}]{BooSmiChe07}%
  \BibitemOpen
  \bibfield  {author} {\bibinfo {author} {\bibfnamefont {A.}~\bibnamefont
  {Boonchun}}, \bibinfo {author} {\bibfnamefont {M.~F.}\ \bibnamefont {Smith}},
  \bibinfo {author} {\bibfnamefont {B.}~\bibnamefont {Cherdhirunkorn}}, \ and\
  \bibinfo {author} {\bibfnamefont {S.}~\bibnamefont {Limpijumnong}},\ }\href
  {\doibase 10.1063/1.2654120} {\bibfield  {journal} {\bibinfo  {journal} {J.
  Appl. Phys.}\ }\textbf {\bibinfo {volume} {101}},\ \bibinfo {pages} {043521}
  (\bibinfo {year} {2007})}\BibitemShut {NoStop}%
\bibitem [{\citenamefont {Eichel}\ \emph {et~al.}(2008)\citenamefont {Eichel},
  \citenamefont {Erhart}, \citenamefont {Tr\"askelin}, \citenamefont {Albe},
  \citenamefont {Kungl},\ and\ \citenamefont {Hoffmann}}]{EicErhTra08}%
  \BibitemOpen
  \bibfield  {author} {\bibinfo {author} {\bibfnamefont {R.-A.}\ \bibnamefont
  {Eichel}}, \bibinfo {author} {\bibfnamefont {P.}~\bibnamefont {Erhart}},
  \bibinfo {author} {\bibfnamefont {P.}~\bibnamefont {Tr\"askelin}}, \bibinfo
  {author} {\bibfnamefont {K.}~\bibnamefont {Albe}}, \bibinfo {author}
  {\bibfnamefont {H.}~\bibnamefont {Kungl}}, \ and\ \bibinfo {author}
  {\bibfnamefont {M.~J.}\ \bibnamefont {Hoffmann}},\ }\href@noop {} {\bibfield
  {journal} {\bibinfo  {journal} {Phys. Rev. Lett.}\ }\textbf {\bibinfo
  {volume} {100}},\ \bibinfo {pages} {095504} (\bibinfo {year}
  {2008})}\BibitemShut {NoStop}%
\bibitem [{\citenamefont {Marton}\ and\ \citenamefont
  {Els\"asser}(2011)}]{MarEls11}%
  \BibitemOpen
  \bibfield  {author} {\bibinfo {author} {\bibfnamefont {P.}~\bibnamefont
  {Marton}}\ and\ \bibinfo {author} {\bibfnamefont {C.}~\bibnamefont
  {Els\"asser}},\ }\href {\doibase 10.1103/PhysRevB.83.020106} {\bibfield
  {journal} {\bibinfo  {journal} {Phys. Rev. B}\ }\textbf {\bibinfo {volume}
  {83}},\ \bibinfo {pages} {020106} (\bibinfo {year} {2011})}\BibitemShut
  {NoStop}%
\bibitem [{\citenamefont {Neumann}\ and\ \citenamefont
  {Arlt}(1987)}]{NeuArl87}%
  \BibitemOpen
  \bibfield  {author} {\bibinfo {author} {\bibfnamefont {H.}~\bibnamefont
  {Neumann}}\ and\ \bibinfo {author} {\bibfnamefont {G.}~\bibnamefont {Arlt}},\
  }\href {\doibase 10.1080/00150198708016950} {\bibfield  {journal} {\bibinfo
  {journal} {Ferroelectrics}\ }\textbf {\bibinfo {volume} {76}},\ \bibinfo
  {pages} {303} (\bibinfo {year} {1987})}\BibitemShut {NoStop}%
\bibitem [{Note1()}]{Note1}%
  \BibitemOpen
  \bibinfo {note} {In Ref.~\protect \rev@citealpnum {ArlNeu88} an alternative
  estimate for the energy differences between different defect dipole
  alignments is given based on dipolar interaction, which leads larger energy
  difference that are closer to the ones obtained by first-principles
  calculations. These values were, however, not employed in said reference to
  actually model aging.}\BibitemShut {Stop}%
\bibitem [{\citenamefont {Jakes}\ \emph {et~al.}(2011)\citenamefont {Jakes},
  \citenamefont {Erdem}, \citenamefont {Eichel}, \citenamefont {Jin},\ and\
  \citenamefont {Damjanovic}}]{JakErdEic11}%
  \BibitemOpen
  \bibfield  {author} {\bibinfo {author} {\bibfnamefont {P.}~\bibnamefont
  {Jakes}}, \bibinfo {author} {\bibfnamefont {E.}~\bibnamefont {Erdem}},
  \bibinfo {author} {\bibfnamefont {R.-A.}\ \bibnamefont {Eichel}}, \bibinfo
  {author} {\bibfnamefont {L.}~\bibnamefont {Jin}}, \ and\ \bibinfo {author}
  {\bibfnamefont {D.}~\bibnamefont {Damjanovic}},\ }\href {\doibase
  10.1063/1.3555465} {\bibfield  {journal} {\bibinfo  {journal} {Appl. Phys.
  Lett.}\ }\textbf {\bibinfo {volume} {98}},\ \bibinfo {pages} {072907}
  (\bibinfo {year} {2011})}\BibitemShut {NoStop}%
\bibitem [{Note2()}]{Note2}%
  \BibitemOpen
  \bibinfo {note} {Since the direct contribution of the electric field to the
  energy landscape is very small (see Sect.~\ref {sect:results_ac}), within our
  model the results for non-oscillating (DC) fields are virtually identical to
  the situation without any external field.}\BibitemShut {Stop}%
\bibitem [{\citenamefont {Granzow}\ \emph {et~al.}(2006)\citenamefont
  {Granzow}, \citenamefont {Suvaci}, \citenamefont {Kungl},\ and\ \citenamefont
  {Hoffmann}}]{GraSuvKun06}%
  \BibitemOpen
  \bibfield  {author} {\bibinfo {author} {\bibfnamefont {T.}~\bibnamefont
  {Granzow}}, \bibinfo {author} {\bibfnamefont {E.}~\bibnamefont {Suvaci}},
  \bibinfo {author} {\bibfnamefont {H.}~\bibnamefont {Kungl}}, \ and\ \bibinfo
  {author} {\bibfnamefont {M.~J.}\ \bibnamefont {Hoffmann}},\ }\href {\doibase
  10.1063/1.2425035} {\bibfield  {journal} {\bibinfo  {journal} {Appl. Phys.
  Lett.}\ }\textbf {\bibinfo {volume} {89}},\ \bibinfo {pages} {262908}
  (\bibinfo {year} {2006})}\BibitemShut {NoStop}%
\bibitem [{\citenamefont {Glaum}\ \emph {et~al.}(2012)\citenamefont {Glaum},
  \citenamefont {Genenko}, \citenamefont {Kungl}, \citenamefont {Schmitt},\
  and\ \citenamefont {Granzow}}]{GlaGenKun12}%
  \BibitemOpen
  \bibfield  {author} {\bibinfo {author} {\bibfnamefont {J.}~\bibnamefont
  {Glaum}}, \bibinfo {author} {\bibfnamefont {Y.~A.}\ \bibnamefont {Genenko}},
  \bibinfo {author} {\bibfnamefont {H.}~\bibnamefont {Kungl}}, \bibinfo
  {author} {\bibfnamefont {L.~A.}\ \bibnamefont {Schmitt}}, \ and\ \bibinfo
  {author} {\bibfnamefont {T.}~\bibnamefont {Granzow}},\ }\href {\doibase
  10.1063/1.4739721} {\bibfield  {journal} {\bibinfo  {journal} {J. Appl.
  Phys.}\ }\textbf {\bibinfo {volume} {112}},\ \bibinfo {pages} {034103}
  (\bibinfo {year} {2012})}\BibitemShut {NoStop}%
\bibitem [{\citenamefont {Ghosez}\ \emph {et~al.}(1999)\citenamefont {Ghosez},
  \citenamefont {Cockayne}, \citenamefont {Waghmare},\ and\ \citenamefont
  {Rabe}}]{GhoCocWag99}%
  \BibitemOpen
  \bibfield  {author} {\bibinfo {author} {\bibfnamefont {P.}~\bibnamefont
  {Ghosez}}, \bibinfo {author} {\bibfnamefont {E.}~\bibnamefont {Cockayne}},
  \bibinfo {author} {\bibfnamefont {U.}~\bibnamefont {Waghmare}}, \ and\
  \bibinfo {author} {\bibfnamefont {K.}~\bibnamefont {Rabe}},\ }\href {\doibase
  10.1103/PhysRevB.60.836} {\bibfield  {journal} {\bibinfo  {journal} {Phys.
  Rev. B}\ }\textbf {\bibinfo {volume} {60}},\ \bibinfo {pages} {836} (\bibinfo
  {year} {1999})}\BibitemShut {NoStop}%
\bibitem [{Note3()}]{Note3}%
  \BibitemOpen
  \bibinfo {note} {In principle, the solution of Eq.~(\ref {eq:model}) can be
  written in terms of the eigenvalues and vectors of $\protect \ensuremath
  \protect \boldsymbol {\protect \mathrm {W}}^{\protect \mathrm {T}}-\protect
  \ensuremath \protect \boldsymbol {\protect \mathrm {V}}$. Since the
  eigenvalues appear in an exponential function, the stability of the solution,
  which can be tested via Eq.~(\ref {eq:ctot}) is highly sensitive to their
  numerical accuracy. In practice, we have therefore resorted to numerical
  solvers that approach Eq.~(\ref {eq:model}) directly.}\BibitemShut {Stop}%
\bibitem [{Note4()}]{Note4}%
  \BibitemOpen
  \bibinfo {note} {Specifically, we used the \protect \texttt {ode15s} solver
  of \protect \textsc {matlab}. \cite {matlab}}\BibitemShut {NoStop}%
\bibitem [{\citenamefont {Kresse}\ and\ \citenamefont
  {Hafner}(1993)}]{KreHaf93}%
  \BibitemOpen
  \bibfield  {author} {\bibinfo {author} {\bibfnamefont {G.}~\bibnamefont
  {Kresse}}\ and\ \bibinfo {author} {\bibfnamefont {J.}~\bibnamefont
  {Hafner}},\ }\href {\doibase 10.1103/PhysRevB.47.558} {\bibfield  {journal}
  {\bibinfo  {journal} {Phys. Rev. B}\ }\textbf {\bibinfo {volume} {47}},\
  \bibinfo {pages} {558} (\bibinfo {year} {1993})}\BibitemShut {NoStop}%
\bibitem [{\citenamefont {Kresse}\ and\ \citenamefont
  {Hafner}(1994)}]{KreHaf94}%
  \BibitemOpen
  \bibfield  {author} {\bibinfo {author} {\bibfnamefont {G.}~\bibnamefont
  {Kresse}}\ and\ \bibinfo {author} {\bibfnamefont {J.}~\bibnamefont
  {Hafner}},\ }\href {\doibase 10.1103/PhysRevB.49.14251} {\bibfield  {journal}
  {\bibinfo  {journal} {Phys. Rev. B}\ }\textbf {\bibinfo {volume} {49}},\
  \bibinfo {pages} {14251} (\bibinfo {year} {1994})}\BibitemShut {NoStop}%
\bibitem [{\citenamefont {Kresse}\ and\ \citenamefont
  {Furthm\"uller}(1996{\natexlab{a}})}]{KreFur96a}%
  \BibitemOpen
  \bibfield  {author} {\bibinfo {author} {\bibfnamefont {G.}~\bibnamefont
  {Kresse}}\ and\ \bibinfo {author} {\bibfnamefont {J.}~\bibnamefont
  {Furthm\"uller}},\ }\href {\doibase 10.1103/PhysRevB.54.11169} {\bibfield
  {journal} {\bibinfo  {journal} {Phys. Rev. B}\ }\textbf {\bibinfo {volume}
  {54}},\ \bibinfo {pages} {11169} (\bibinfo {year}
  {1996}{\natexlab{a}})}\BibitemShut {NoStop}%
\bibitem [{\citenamefont {Kresse}\ and\ \citenamefont
  {Furthm\"uller}(1996{\natexlab{b}})}]{KreFur96b}%
  \BibitemOpen
  \bibfield  {author} {\bibinfo {author} {\bibfnamefont {G.}~\bibnamefont
  {Kresse}}\ and\ \bibinfo {author} {\bibfnamefont {J.}~\bibnamefont
  {Furthm\"uller}},\ }\href {\doibase 10.1016/0927-0256(96)00008-0} {\bibfield
  {journal} {\bibinfo  {journal} {Comput. Mater. Sci.}\ }\textbf {\bibinfo
  {volume} {6}},\ \bibinfo {pages} {15} (\bibinfo {year}
  {1996}{\natexlab{b}})}\BibitemShut {NoStop}%
\bibitem [{\citenamefont {Bl\"ochl}(1994)}]{Blo94}%
  \BibitemOpen
  \bibfield  {author} {\bibinfo {author} {\bibfnamefont {P.~E.}\ \bibnamefont
  {Bl\"ochl}},\ }\href {\doibase 10.1103/PhysRevB.54.11169} {\bibfield
  {journal} {\bibinfo  {journal} {Phys. Rev. B}\ }\textbf {\bibinfo {volume}
  {50}},\ \bibinfo {pages} {17953 } (\bibinfo {year} {1994})}\BibitemShut
  {NoStop}%
\bibitem [{\citenamefont {Kresse}\ and\ \citenamefont
  {Joubert}(1999)}]{KreJou99}%
  \BibitemOpen
  \bibfield  {author} {\bibinfo {author} {\bibfnamefont {G.}~\bibnamefont
  {Kresse}}\ and\ \bibinfo {author} {\bibfnamefont {D.}~\bibnamefont
  {Joubert}},\ }\href {\doibase 10.1103/PhysRevB.59.1758} {\bibfield  {journal}
  {\bibinfo  {journal} {Phys. Rev. B}\ }\textbf {\bibinfo {volume} {59}},\
  \bibinfo {pages} {1758 } (\bibinfo {year} {1999})}\BibitemShut {NoStop}%
\bibitem [{\citenamefont {Ceperley}\ and\ \citenamefont
  {Alder}(1980)}]{CepAld80}%
  \BibitemOpen
  \bibfield  {author} {\bibinfo {author} {\bibfnamefont {D.~M.}\ \bibnamefont
  {Ceperley}}\ and\ \bibinfo {author} {\bibfnamefont {B.~J.}\ \bibnamefont
  {Alder}},\ }\href {\doibase 10.1103/PhysRevLett.45.566} {\bibfield  {journal}
  {\bibinfo  {journal} {Phys. Rev. Lett.}\ }\textbf {\bibinfo {volume} {45}},\
  \bibinfo {pages} {566} (\bibinfo {year} {1980})}\BibitemShut {NoStop}%
\bibitem [{\citenamefont {Perdew}\ and\ \citenamefont
  {Zunger}(1981)}]{PerZun81}%
  \BibitemOpen
  \bibfield  {author} {\bibinfo {author} {\bibfnamefont {J.~P.}\ \bibnamefont
  {Perdew}}\ and\ \bibinfo {author} {\bibfnamefont {A.}~\bibnamefont
  {Zunger}},\ }\href {\doibase 10.1103/PhysRevB.23.5048} {\bibfield  {journal}
  {\bibinfo  {journal} {Phys. Rev. B}\ }\textbf {\bibinfo {volume} {23}},\
  \bibinfo {pages} {5048} (\bibinfo {year} {1981})}\BibitemShut {NoStop}%
\bibitem [{\citenamefont {Glazer}\ and\ \citenamefont
  {Mabud}(1978)}]{GlaMab78b}%
  \BibitemOpen
  \bibfield  {author} {\bibinfo {author} {\bibfnamefont {A.~M.}\ \bibnamefont
  {Glazer}}\ and\ \bibinfo {author} {\bibfnamefont {S.~A.}\ \bibnamefont
  {Mabud}},\ }\href {\doibase 10.1107/S0567740878004938} {\bibfield  {journal}
  {\bibinfo  {journal} {Acta Crystallogr. B}\ }\textbf {\bibinfo {volume}
  {34}},\ \bibinfo {pages} {1065} (\bibinfo {year} {1978})}\BibitemShut
  {NoStop}%
\bibitem [{\citenamefont {Robertson}\ and\ \citenamefont
  {Chen}(1999)}]{RobChe99}%
  \BibitemOpen
  \bibfield  {author} {\bibinfo {author} {\bibfnamefont {J.}~\bibnamefont
  {Robertson}}\ and\ \bibinfo {author} {\bibfnamefont {C.~W.}\ \bibnamefont
  {Chen}},\ }\href {\doibase 10.1063/1.123476} {\bibfield  {journal} {\bibinfo
  {journal} {Appl. Phys. Lett.}\ }\textbf {\bibinfo {volume} {74}},\ \bibinfo
  {pages} {1168} (\bibinfo {year} {1999})}\BibitemShut {NoStop}%
\bibitem [{\citenamefont {Erhart}\ and\ \citenamefont
  {Albe}(2006)}]{ErhAlb06a}%
  \BibitemOpen
  \bibfield  {author} {\bibinfo {author} {\bibfnamefont {P.}~\bibnamefont
  {Erhart}}\ and\ \bibinfo {author} {\bibfnamefont {K.}~\bibnamefont {Albe}},\
  }\href {\doibase 10.1103/PhysRevB.73.115207} {\bibfield  {journal} {\bibinfo
  {journal} {Phys. Rev. B}\ }\textbf {\bibinfo {volume} {73}},\ \bibinfo
  {pages} {115207} (\bibinfo {year} {2006})}\BibitemShut {NoStop}%
\bibitem [{\citenamefont {Henkelman}\ \emph
  {et~al.}(2000{\natexlab{a}})\citenamefont {Henkelman}, \citenamefont
  {J\'ohannesson},\ and\ \citenamefont {J\'onsson}}]{HenJohJon00}%
  \BibitemOpen
  \bibfield  {author} {\bibinfo {author} {\bibfnamefont {G.}~\bibnamefont
  {Henkelman}}, \bibinfo {author} {\bibfnamefont {G.}~\bibnamefont
  {J\'ohannesson}}, \ and\ \bibinfo {author} {\bibfnamefont {H.}~\bibnamefont
  {J\'onsson}},\ }\enquote {\bibinfo {title} {Methods for finding saddlepoints
  and minimum energy paths},}\ \ (\bibinfo  {publisher} {Kluwer Academic},\
  \bibinfo {address} {Dordrecht},\ \bibinfo {year} {2000})\ \bibinfo {note}
  {\textup{in} {\em Progress on theoretical chemistry and physics}, \textup{p.
  269}}\BibitemShut {NoStop}%
\bibitem [{\citenamefont {Henkelman}\ \emph
  {et~al.}(2000{\natexlab{b}})\citenamefont {Henkelman}, \citenamefont
  {Uberuaga},\ and\ \citenamefont {J\'onsson}}]{HenUbeJon00}%
  \BibitemOpen
  \bibfield  {author} {\bibinfo {author} {\bibfnamefont {G.}~\bibnamefont
  {Henkelman}}, \bibinfo {author} {\bibfnamefont {B.~P.}\ \bibnamefont
  {Uberuaga}}, \ and\ \bibinfo {author} {\bibfnamefont {H.}~\bibnamefont
  {J\'onsson}},\ }\href {\doibase 10.1063/1.1329672} {\bibfield  {journal}
  {\bibinfo  {journal} {J. Chem. Phys.}\ }\textbf {\bibinfo {volume} {113}},\
  \bibinfo {pages} {9901} (\bibinfo {year} {2000}{\natexlab{b}})}\BibitemShut
  {NoStop}%
\bibitem [{\citenamefont {Park}\ and\ \citenamefont {Chadi}(1998)}]{ParCha98}%
  \BibitemOpen
  \bibfield  {author} {\bibinfo {author} {\bibfnamefont {C.~H.}\ \bibnamefont
  {Park}}\ and\ \bibinfo {author} {\bibfnamefont {D.~J.}\ \bibnamefont
  {Chadi}},\ }\href {\doibase 10.1103/PhysRevB.57.R13961} {\bibfield  {journal}
  {\bibinfo  {journal} {Phys. Rev. B}\ }\textbf {\bibinfo {volume} {57}},\
  \bibinfo {pages} {R13961} (\bibinfo {year} {1998})}\BibitemShut {NoStop}%
\bibitem [{Note5()}]{Note5}%
  \BibitemOpen
  \bibinfo {note} {Park and Chady \cite {ParCha98} discuss two different
  configurations for the $ab$-site vacancy, $V_{ab}^{sw}$ and $V_{ab}^{ud}$,
  but the former one seems to be always lower in energy and therefore prevails.
  We are thus left with only two different types vacancies, $V_c$ and $V_{ab}$.
  Also compare the discussion in Ref.~\protect \rev@citealpnum
  {ErhEicTra07}.}\BibitemShut {Stop}%
\bibitem [{\citenamefont {Park}(2003)}]{Par03}%
  \BibitemOpen
  \bibfield  {author} {\bibinfo {author} {\bibfnamefont {C.~H.}\ \bibnamefont
  {Park}},\ }\href@noop {} {\bibfield  {journal} {\bibinfo  {journal} {J.
  Korean Phys. Soc.}\ }\textbf {\bibinfo {volume} {42}},\ \bibinfo {pages}
  {S1420 } (\bibinfo {year} {2003})}\BibitemShut {NoStop}%
\bibitem [{\citenamefont {Erhart}\ and\ \citenamefont {Albe}(2007)}]{ErhAlb07}%
  \BibitemOpen
  \bibfield  {author} {\bibinfo {author} {\bibfnamefont {P.}~\bibnamefont
  {Erhart}}\ and\ \bibinfo {author} {\bibfnamefont {K.}~\bibnamefont {Albe}},\
  }\href {\doibase 10.1063/1.2801011} {\bibfield  {journal} {\bibinfo
  {journal} {J. Appl. Phys.}\ }\textbf {\bibinfo {volume} {102}},\ \bibinfo
  {pages} {084111} (\bibinfo {year} {2007})}\BibitemShut {NoStop}%
\bibitem [{\citenamefont {Gottschalk}\ \emph {et~al.}(2008)\citenamefont
  {Gottschalk}, \citenamefont {Hahn}, \citenamefont {Flege},\ and\
  \citenamefont {Balogh}}]{GotHahFle08}%
  \BibitemOpen
  \bibfield  {author} {\bibinfo {author} {\bibfnamefont {S.}~\bibnamefont
  {Gottschalk}}, \bibinfo {author} {\bibfnamefont {H.}~\bibnamefont {Hahn}},
  \bibinfo {author} {\bibfnamefont {S.}~\bibnamefont {Flege}}, \ and\ \bibinfo
  {author} {\bibfnamefont {A.~G.}\ \bibnamefont {Balogh}},\ }\href {\doibase
  10.1063/1.2988902} {\bibfield  {journal} {\bibinfo  {journal} {J. Appl.
  Phys.}\ }\textbf {\bibinfo {volume} {104}},\ \bibinfo {pages} {114106}
  (\bibinfo {year} {2008})}\BibitemShut {NoStop}%
\bibitem [{Note6()}]{Note6}%
  \BibitemOpen
  \bibinfo {note} {At finite temperature thermodynamically a certain fraction
  of vacancies will not occupy the lowest energy configuration. Their number,
  however, is very small (and indistinguishable from zero on the scale of
  Fig.~\ref {fig:Cu_dc}(a)), since it is determined by the Boltzmann factor
  $\protect \qopname \relax o{exp}(-\Delta E/k_B T)$ where $\Delta E$ is the
  energy difference between the ground state configuration, $M_{\protect \text
  {Ti}}-V_{c1}$, and the configuration in question.}\BibitemShut {Stop}%
\bibitem [{\citenamefont {Noheda}\ \emph {et~al.}(1995)\citenamefont {Noheda},
  \citenamefont {Cereceda}, \citenamefont {Iglesias}, \citenamefont {Lifante},
  \citenamefont {Gonzalo}, \citenamefont {Chen},\ and\ \citenamefont
  {Wang}}]{NohCerIgl95}%
  \BibitemOpen
  \bibfield  {author} {\bibinfo {author} {\bibfnamefont {B.}~\bibnamefont
  {Noheda}}, \bibinfo {author} {\bibfnamefont {N.}~\bibnamefont {Cereceda}},
  \bibinfo {author} {\bibfnamefont {T.}~\bibnamefont {Iglesias}}, \bibinfo
  {author} {\bibfnamefont {G.}~\bibnamefont {Lifante}}, \bibinfo {author}
  {\bibfnamefont {J.~A.}\ \bibnamefont {Gonzalo}}, \bibinfo {author}
  {\bibfnamefont {H.~T.}\ \bibnamefont {Chen}}, \ and\ \bibinfo {author}
  {\bibfnamefont {Y.~L.}\ \bibnamefont {Wang}},\ }\href {\doibase
  10.1103/PhysRevB.51.16388} {\bibfield  {journal} {\bibinfo  {journal} {Phys.
  Rev. B}\ }\textbf {\bibinfo {volume} {51}},\ \bibinfo {pages} {16388}
  (\bibinfo {year} {1995})}\BibitemShut {NoStop}%
\bibitem [{Note7()}]{Note7}%
  \BibitemOpen
  \bibinfo {note} {The Curie temperatures of PbTiO$_3$ and PbZrO$_3$ are 720
  and 460\protect \tmspace +\thinmuskip {.1667em}K, respectively, \cite
  {NohCerIgl95} and the concentration dependent Curie temperature of PZT is
  bounded by these values.}\BibitemShut {Stop}%
\bibitem [{\citenamefont {Meyer}\ and\ \citenamefont
  {Vanderbilt}(2002)}]{MeyVan02}%
  \BibitemOpen
  \bibfield  {author} {\bibinfo {author} {\bibfnamefont {B.}~\bibnamefont
  {Meyer}}\ and\ \bibinfo {author} {\bibfnamefont {D.}~\bibnamefont
  {Vanderbilt}},\ }\href {\doibase 10.1103/PhysRevB.65.104111} {\bibfield
  {journal} {\bibinfo  {journal} {Phys. Rev. B}\ }\textbf {\bibinfo {volume}
  {65}},\ \bibinfo {pages} {104111} (\bibinfo {year} {2002})}\BibitemShut
  {NoStop}%
\bibitem [{\citenamefont {Cockayne}\ and\ \citenamefont
  {Burton}(2004)}]{CocBur04}%
  \BibitemOpen
  \bibfield  {author} {\bibinfo {author} {\bibfnamefont {E.}~\bibnamefont
  {Cockayne}}\ and\ \bibinfo {author} {\bibfnamefont {B.~P.}\ \bibnamefont
  {Burton}},\ }\href {\doibase 10.1103/PhysRevB.69.144116} {\bibfield
  {journal} {\bibinfo  {journal} {Phys. Rev. B}\ }\textbf {\bibinfo {volume}
  {69}},\ \bibinfo {pages} {144116} (\bibinfo {year} {2004})}\BibitemShut
  {NoStop}%
\bibitem [{\citenamefont {MATLAB}(2010)}]{matlab}%
  \BibitemOpen
  \bibfield  {author} {\bibinfo {author} {\bibnamefont {MATLAB}},\ }\href@noop
  {} {\emph {\bibinfo {title} {version 7.10.0 (R2010a)}}}\ (\bibinfo
  {publisher} {The MathWorks Inc.},\ \bibinfo {address} {Natick,
  Massachusetts},\ \bibinfo {year} {2010})\BibitemShut {NoStop}%
\end{thebibliography}
\end{document}